\documentclass[prc,twocolumn,superscriptaddress,showpacs,twoside,floatfix]
{revtex4-1}

\usepackage{amssymb,epsfig}
\usepackage{color}

\begin{document}

\title{Deep excursion beyond the proton dripline. I. Argon and chlorine isotope chains}

\author{I.~Mukha}
\affiliation{GSI Helmholtzzentrum  f\"{u}r Schwerionenforschung GmbH, 64291
Darmstadt, Germany}

\author{L.V.~Grigorenko}
\affiliation{Flerov Laboratory of Nuclear Reactions, JINR,  141980 Dubna, Russia}
\affiliation{National Research Nuclear University ``MEPhI'',
115409 Moscow, Russia}
\affiliation{National Research Centre ``Kurchatov Institute'', Kurchatov
sq.\ 1, 123182 Moscow, Russia}

\author{D.~Kostyleva}
\email{D.Kostyleva@gsi.de}
\affiliation{II.Physikalisches Institut, Justus-Liebig-Universit\"at, 35392 Gie{\ss}en, Germany}
\affiliation{GSI Helmholtzzentrum  f\"{u}r Schwerionenforschung, 64291
Darmstadt, Germany}

\author{L.~Acosta}
\affiliation{INFN, Laboratori Nazionali del Sud,
Via S.~Sof\'ia, 95123 Catania, Italy}
\affiliation{Instituto de F\'isica, Universidad Nacional Aut\'onoma de M\'exico, M\'exico, D.F.\ 01000, Mexico}

\author{E.~Casarejos}
\affiliation{University of  Vigo, 36310 Vigo, Spain}

\author{A.A.~Ciemny}
\affiliation{Faculty of Physics, University of Warsaw, 02-093 Warszawa, Poland}

\author{W.~Dominik}
\affiliation{Faculty of Physics, University of Warsaw, 02-093 Warszawa, Poland}

\author{J.A.~Due\~nas}
\affiliation{Depto. de Ingenieria Electrica y Centro de Estudios Avanzados en Fisica, Matem\'{a}ticas y Computaci\'{o}n, Universidad de Huelva, 21071 Huelva, Spain}

\author{V.~Dunin}
\affiliation{Veksler and Baldin Laboratory of High Energy Physics, JINR, 141980 Dubna,
Russia}

\author{J.~M.~Espino}
\affiliation{Department of Atomic, Molecular and Nuclear Physics, University of Seville, 41012 Seville, Spain}

\author{A.~Estrad\'{e}}
\affiliation{University of Edinburgh, EH1 1HT Edinburgh, United Kingdom}

\author{F.~Farinon}
\affiliation{GSI Helmholtzzentrum  f\"{u}r Schwerionenforschung, 64291 Darmstadt, Germany}

\author{A.~Fomichev}
\affiliation{Flerov Laboratory of Nuclear Reactions, JINR,  141980 Dubna, Russia}

\author{H.~Geissel}
\affiliation{GSI Helmholtzzentrum  f\"{u}r Schwerionenforschung, 64291 Darmstadt, Germany}
\affiliation{II.Physikalisches Institut, Justus-Liebig-Universit\"at, 35392 Gie{\ss}en, Germany}

\author{A.~Gorshkov}
\affiliation{Flerov Laboratory of Nuclear Reactions, JINR,  141980 Dubna, Russia}

\author{Z.~Janas}
 \affiliation{Faculty of Physics,  University of Warsaw, 02-093 Warszawa,
 Poland}

\author{G.~Kami\'{n}ski}
\affiliation{Heavy Ion Laboratory, University of Warsaw, 02-093 Warszawa,
Poland}
\affiliation{Flerov Laboratory of Nuclear Reactions, JINR,  141980 Dubna,
Russia}

\author{O.~Kiselev}
\affiliation{GSI Helmholtzzentrum  f\"{u}r Schwerionenforschung, 64291 Darmstadt, Germany}

\author{R.~Kn\"{o}bel}
\affiliation{GSI Helmholtzzentrum  f\"{u}r Schwerionenforschung, 64291 Darmstadt, Germany}
\affiliation{II.Physikalisches Institut, Justus-Liebig-Universit\"at, 35392 Gie{\ss}en, Germany}

\author{S.~Krupko}
\affiliation{Flerov Laboratory of Nuclear Reactions, JINR,  141980 Dubna, Russia}

\author{M.~Kuich}
\affiliation{Faculty of Physics, Warsaw University of Technology, 00-662 Warszawa, Poland}
 \affiliation{Faculty of Physics,  University of Warsaw, 02-093 Warszawa,
 Poland}

\author{Yu.A.~Litvinov}
\affiliation{GSI Helmholtzzentrum  f\"{u}r Schwerionenforschung, 64291 Darmstadt, Germany}

\author{G.~Marquinez-Dur\'{a}n}
\affiliation{Department of Applied Physics, University of Huelva, 21071 Huelva, Spain}

\author{I.~Martel}
\affiliation{Department of Applied Physics, University of Huelva, 21071 Huelva, Spain}

\author{C.~Mazzocchi}
\affiliation{Faculty of Physics, University of Warsaw, 02-093 Warszawa, Poland}

\author{C.~Nociforo}
\affiliation{GSI Helmholtzzentrum  f\"{u}r Schwerionenforschung, 64291 Darmstadt, Germany}

\author{A.~K.~Ord\'{u}z}
\affiliation{Department of Applied Physics, University of Huelva, 21071 Huelva, Spain}

\author{M.~Pf\"{u}tzner}
\affiliation{Faculty of Physics, University of Warsaw, 02-093 Warszawa, Poland}
\affiliation{GSI Helmholtzzentrum  f\"{u}r Schwerionenforschung, 64291 Darmstadt, Germany}

\author{S.~Pietri}
\affiliation{GSI Helmholtzzentrum  f\"{u}r Schwerionenforschung, 64291 Darmstadt, Germany}

\author{M.~Pomorski}
 \affiliation{Faculty of Physics,  University of Warsaw, 02-093 Warszawa,
 Poland}

\author{A.~Prochazka}
\affiliation{GSI Helmholtzzentrum  f\"{u}r Schwerionenforschung, 64291 Darmstadt, Germany}

\author{S.~Rymzhanova}
\affiliation{Flerov Laboratory of Nuclear Reactions, JINR,  141980 Dubna, Russia}

\author{A.M.~S\'{a}nchez-Ben\'{i}tez}
\affiliation{Centro de Estudios Avanzados en F\'{i}sica, Matem\'{a}ticas y
Computaci\'{o}n (CEAFMC), Department of Integrated Sciences, University of
Huelva,  21071 Huelva, Spain}

\author{C.~Scheidenberger}
\affiliation{GSI Helmholtzzentrum  f\"{u}r Schwerionenforschung, 64291 Darmstadt, Germany}
\affiliation{II.Physikalisches Institut, Justus-Liebig-Universit\"at, 35392 Gie{\ss}en, Germany}

\author{P.~Sharov}
\affiliation{Flerov Laboratory of Nuclear Reactions, JINR,  141980 Dubna, Russia}

\author{H.~Simon}
\affiliation{GSI Helmholtzzentrum  f\"{u}r Schwerionenforschung, 64291 Darmstadt, Germany}

\author{B.~Sitar}
\affiliation{Faculty of Mathematics and Physics, Comenius University, 84248 Bratislava,
Slovakia}

\author{R.~Slepnev}
\affiliation{Flerov Laboratory of Nuclear Reactions, JINR,  141980 Dubna, Russia}

\author{M.~Stanoiu}
\affiliation{IFIN-HH, Post Office Box MG-6, Bucharest, Romania}

\author{P.~Strmen}
\affiliation{Faculty of Mathematics and Physics, Comenius University, 84248 Bratislava,
Slovakia}

\author{I.~Szarka}
\affiliation{Faculty of Mathematics and Physics, Comenius University, 84248 Bratislava,
Slovakia}

\author{M.~Takechi}
\affiliation{GSI Helmholtzzentrum  f\"{u}r Schwerionenforschung, 64291 Darmstadt, Germany}

\author{Y.K.~Tanaka}
\affiliation{GSI Helmholtzzentrum  f\"{u}r Schwerionenforschung, 64291 Darmstadt, Germany}
\affiliation{University of Tokyo, 113-0033 Tokyo, Japan}

\author{H.~Weick}
\affiliation{GSI Helmholtzzentrum  f\"{u}r Schwerionenforschung, 64291 Darmstadt, Germany}

\author{M.~Winkler}
\affiliation{GSI Helmholtzzentrum  f\"{u}r Schwerionenforschung, 64291 Darmstadt, Germany}

\author{J.S.~Winfield}
\affiliation{GSI Helmholtzzentrum  f\"{u}r Schwerionenforschung, 64291 Darmstadt, Germany}

\author{X.~Xu}
\affiliation{School of Physics and Nuclear Energy Engineering, Beihang University, 100191 Beijing, China}
\affiliation{II.Physikalisches Institut, Justus-Liebig-Universit\"at, 35392 Gie{\ss}en, Germany}
\affiliation{GSI Helmholtzzentrum  f\"{u}r Schwerionenforschung, 64291 Darmstadt, Germany}

\author{M.V.~Zhukov}
\affiliation{Department of Physics, Chalmers University of Technology, S-41296 G\"oteborg, Sweden}

\collaboration{ for the Super-FRS Experiment Collaboration}

\date{\today. {\tt File: ar-cl-excur-f1-9.tex }}

\begin{abstract}
The proton-unbound argon and chlorine isotopes have been studied by measuring trajectories of their decay-in-flight products by using a tracking technique with micro-strip detectors. The proton ($1p$) and two-proton ($2p$) emission processes have been detected in the measured angular correlations  ``heavy-fragment''+$p$ and ``heavy-fragment''+$p$+$p$, respectively. The ground states of the previously unknown isotopes $^{30}$Cl and $^{28}$Cl have been observed for the first time, providing the $1p$ separation energies $S_p$ of $-0.48(2)$ and $-1.60(8)$ MeV, respectively. The relevant systematics of $1p$ and $2p$ separation energies have been studied theoretically in the core+$p$ and core+$p$+$p$ cluster models. The first-time observed excited states of $^{31}$Ar allow to infer the $2p$-separation energy $S_{2p}$ of 6(34)~keV for its ground state. The first-time observed state in  $^{29}$Ar with $S_{2p} = -5.50(18)$~MeV can be identified either as a ground or an excited state according to different systematics.
\end{abstract}


\maketitle


\section{Introduction}


The location of the driplines --- the borderlines separating particle-stable and particle-unstable isotopes --- is one of the fundamental questions of nuclear science. The unbound states with small decay energy can have lifetimes which are long enough to be treated as quasistationary states. Thus they may be considered as stationary states in many theoretical applications. This naturally leads us to the question: what are the \emph{limits of nuclear structure existence}? In other words, how far beyond the driplines the nuclear structure phenomena fade and are completely replaced by the continuum dynamics? This question represents a motivation for studies of nuclear systems far beyond the driplines.

The proton and neutron driplines have been accessed for nuclides in broad ranges of $Z$ (number of protons) and $N$ (number of neutrons) of the nuclear chart. However, even in these regions the information about the nearest to the dripline unbound isotopes is scarce and often missing. Thus the fundamental question about the limits of the nuclear structure existence remains poorly investigated. For example, if we consider the proton dripline within $Z \leq 20$  ($p$- and $sd$-shell nuclei), the most extensively investigated case in that region is the fluoride isotope chain. Here our knowledge extends three mass units beyond the proton dripline: the $^{15}$F and $^{16}$F  nuclides are well studied, and considerable spectroscopic information is available now for $^{14}$F \cite{Goldberg:2010} in addition.

This paper continues our analysis of the data on reactions with a relativistic $^{31}$Ar beam populating particle-unstable states  \cite{Mukha:2015,Golubkova:2016,Xu:2018}. The article \cite{Mukha:2015} was focused on $^{30}$Ar and $^{29}$Cl isotopes which were reported for the first time. It was also found that the decay mechanism of $^{30}$Ar is likely to belong to a transition region between true $2p$ and sequential $2p$ decay mechanisms. Such a ``transition regime'' exhibits strong sensitivity of observed kinematic variables to the values of parameters defining the decay mechanism: $2p$-decay energy $E_T$, ground state (g.s.) resonance energy in the core+$p$ subsystem $E_r$, and its width $\Gamma_r$. The practical implementations of this fact, including opportunity of a precise determination of $\Gamma_r$ from the $2p$ correlation data, were recently elaborated in Ref.\ \cite{Golubkova:2016}. In paper \cite{Xu:2018} a detailed consideration of the data from \cite{Mukha:2015} was given.

In present work we report on the byproduct data of the same experiment which resulted in Refs.\ \cite{Mukha:2015,Golubkova:2016,Xu:2018}, which include observation of $^{28}$Cl and $^{30}$Cl ground states and several (presumably excited) states in $^{29}$Ar and $^{31}$Ar. In order to clarify the situation with the observed states, we have performed systematic studies of separation energies in the chlorine and argon isotope chains. The depth of the performed ``excursion beyond the proton dripline'' in the argon and chlorine isotope chains is similar now in extent to that for the fluorine isotope chain, the best-studied case in the whole $Z\leq 20$ nuclei region.


\section{Experiment}

\begin{figure}[tb]
\begin{center}
\includegraphics[width=0.49\textwidth]{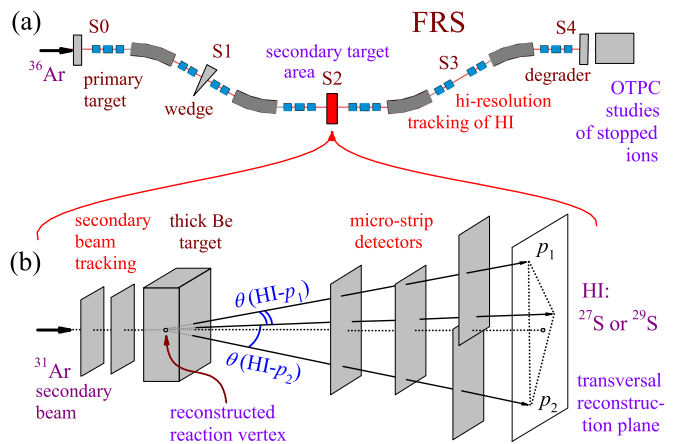}
\end{center}
\caption{  Sketch of the experiment at the FRS fragment separator. (a) General layout, see details in \cite{Xu:2018}. Primary $^{36}$Ar beam hits a target at S0, and a secondary beam of $^{31}$Ar is selected and focused on the middle focal plane S2 by using a wedge degrader at S1. Nuclei of interest $^{31}\text{Ar}^*$, $^{29}$Ar are produced in the secondary target at S2 in in-elastic scattering and two-neutron knock-out reactions. Heavy-ion decay products $^{29,27}$S are identified by their energy-loss and time-of-flight as well as the magnetic rigidity by using the standard beam detectors at S2 and S4. At S4, the studies of stopped ions can be performed by using Optical Time Projection Chamber OTPC. (b) The detector setup downstream of the secondary target at S2, where the trajectories of an incoming $^{31}$Ar ions, the decay products $^{27,29}$S and the protons (i.e., $p_1, p_2$) are measured.}
\label{fig:scheme}
\end{figure}

In the experiment, described in detail in Refs.\ \cite{Mukha:2015,Xu:2018}, the $^{31}$Ar beam was obtained by the fragmentation of a primary 885 \emph{A}MeV $^{36}$Ar beam at the SIS-FRS facility at GSI (Germany). The prime objective of the experiment was study of $2p$ decays of $^{30}$Ar isotopes. The scheme of the measurements is shown in Fig.~\ref{fig:scheme} (a). We briefly repeat the general description of the experiment and the detector performance given in Ref.\ \cite{Xu:2018} in detail.

The FRS was operated with an ion-optical settings in a separator-spectrometer mode, when the first half of the FRS was set for separation and focusing of the radioactive beams on a secondary target in the middle of the FRS, and the second half of FRS was set for detection of heavy-ion decay products. The 620 AMeV $^{31}$Ar ions with the intensity of 50 ions $\text{s}^{-1}$ were transported by the first half of the FRS in order to bombard a $^9$Be secondary target located at the middle focal plane S2. At the first focal plane S1 of the FRS, an aluminum wedge degrader was installed in order to achieve an achromatic focusing of $^{31}$Ar at the secondary target. In the previously reported data \cite{Xu:2018} the $^{30}$Ar nuclei were produced via one-neutron ($-1n$) knockout from the $^{31}$Ar ions. The decay products of unbound $^{30}$Ar were tracked by a double-sided silicon micro-strip detector array placed just downstream of the secondary target, see Fig.~\ref{fig:scheme} (b). The projectile-like particles, outgoing from the secondary target, were analyzed by the second half of the FRS, which was operated as a magnetic spectrometer. The magnet settings between the S2 and S4 focal planes were tuned for the transmission of the targeted heavy ion (HI) fragments (e.g., $^{28}$S) down to S4, the last focal plane. In addition to $^{30}$Ar, the studies of decay properties of the stopped $^{31}$Ar ions were performed by using the OTPC detector at S4 \cite{Lis:2015}.

A double-sided silicon micro-strip detector (DSSD) array consisted out of four large-area DSSDs \cite{Stanoiu:2008} was employed to measure hit coordinates of the two protons and the recoil heavy ions, resulting from the in-flight decays of the studied $2p$ precursors. The high-precision position measurement by DSSDs allowed for reconstruction of all fragment trajectories, which let us to derive the decay vertex together with an angular HI-$p$ and HI-$p$-$p$ correlations. For example, the trajectories of measured $^{28}$S+$p$+$p$ coincidences were a basis for the analysis and the concluded spectroscopic information on $^{30}$Ar \cite{Xu:2018}.

\begin{figure}[t!]
\begin{center}
\includegraphics[width=0.81\linewidth]{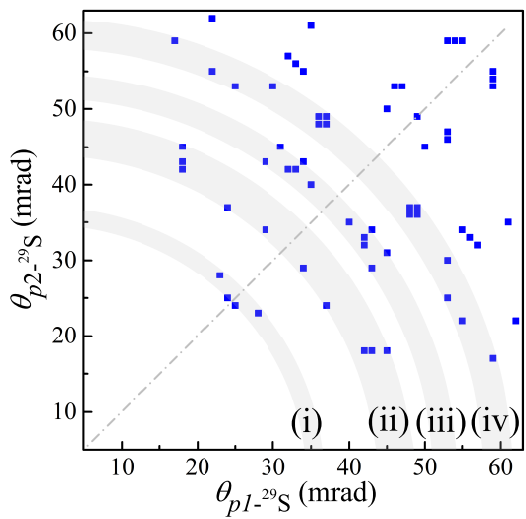}
\end{center}
%
\caption{
  Angular correlations $\theta_{p_1\text{-}^{29}\text{S}}-\theta_{p_2\text{-}^{29}\text{S}}$  (in mradians) produced from the measured  $^{29}$S+$p$+$p$ coincidence indicating $2p$ decays of $^{31}$Ar excited states. The shaded arcs labeled by the Roman numerals point to four selected areas where the $2p$-decay events have the same decay energy. The data symmetry respective proton permutations is illustrated by the $45^\circ$-tilted diagonal dash-dot line.}
\label{fig:Ang1_Ang2_S29pp}
\end{figure}

\begin{figure}[tb]
\includegraphics[width=0.89\linewidth]{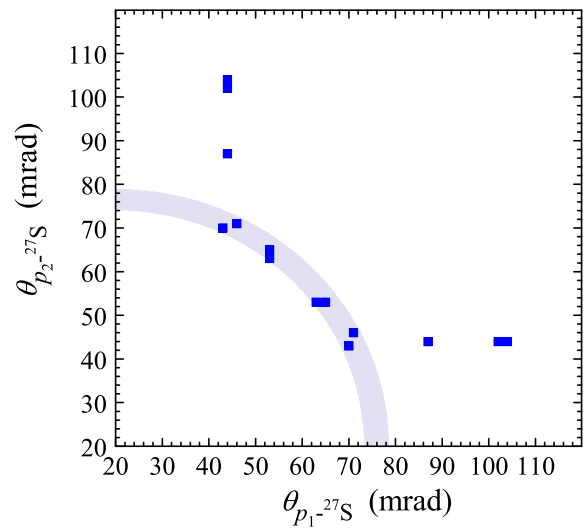}
%
\caption{  Angular correlations $\theta_{p_1\text{-}^{27}\text{S}}-\theta_{p_2\text{-}^{27}\text{S}}$ (in mradians) produced from the measured  $^{27}$S+$p$+$p$ coincidence. The shaded arc where the $2p$-decay events have the same decay energy indicates on a $^{29}$Ar state.
}
\label{fig:Ang1_Ang2_S27pp}
\end{figure}

However, a number of by-product results were obtained in a similar way from the data recorded in the same experiment. Namely,  excited states of $^{31}$Ar were populated by various inelastic mechanisms, and $^{29}$Ar spectrum was populated in two-neutron ($-2n$) knockout reaction. The unbound $^{31}$Ar and $^{29}$Ar states were detected in triple $^{29}$S+$p$+$p$ and $^{27}$S+$p$+$p$ coincidences, respectively, see  the respective angular correlation plots in Figs.\ \ref{fig:Ang1_Ang2_S29pp} and \ref{fig:Ang1_Ang2_S27pp}. The relative angles there and everywhere below are presented in milliradian units (mrad). Also the states of $^{28}$Cl and $^{30}$Cl can be populated both in the fragmentation of $^{31}$Ar and as the result of proton emission from the corresponding $^{31,29}$Ar isotopes. These mechanisms have lower cross sections, and the obtained results have less statistics, see the respective angular $p$-HI correlation plots in Figs.\ \ref{fig:exp-31ar} and \ref{fig:exp-29ar}. In spite of poor statistics with few events registered, we have obtained several nuclear-structure conclusions from the experimental data.

%
%

\subsection{How nuclear-structure information can be obtained from proton-ion angular correlations}

Before the data analysis presentation, we remind the reader, how nuclear-structure
information concerning the nuclei involved in a $1p$ or $2p$ decay can
be obtained by measuring only the trajectories of the decay
products, without measuring their kinetic energies. This approach has been successfully tested in analyses of $1p$, $2p$ decays of the known states in $^{19}$Na, $^{16}$Ne and has been described in details in Ref.\ \cite{Mukha:2010}.

%
\begin{figure}[tb]
\begin{center}
\includegraphics[width=0.99\linewidth]{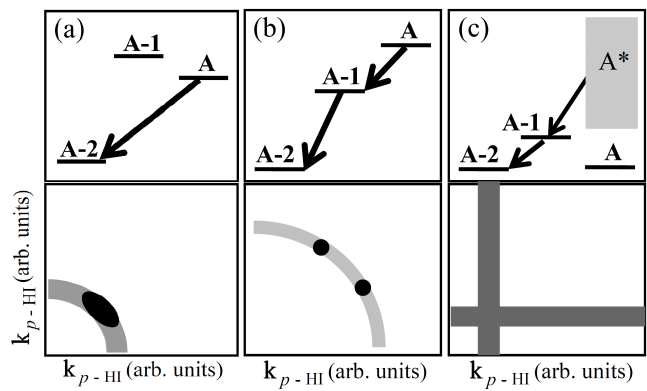}
\end{center}
\caption{Sketches of decay schemes expected for three simple mechanisms of $2p$ decay from a $2p$-precursor with mass number $A$ to a daughter nucleus with $A-2$ (illustrated in the upper panels). The respective cartoons of transverse
momentum correlations $k_{p_1\text{-HI}}-k_{p_2\text{-HI}}$ are shown in the lower panels: (a)  prompt or direct 2\emph{p} emission (three-body decay); (b) sequential emission of protons via a narrow intermediate state in nucleus $A-1$; (c) de-excitation of broad continuum states in the nucleus $A$ to a single low-lying resonance in the intermediate nucleus with $A-1$.}
\label{fig:kinem_a_b_c}
\end{figure}
%

For the discussion of $2p$ decays given below, let us consider three different mechanisms. These cases are illustrated in Fig.~\ref{fig:kinem_a_b_c}. The upper panels schematically show the nuclear states involved in $2p$ decay of nucleus with mass number $A$, the lower panels show the corresponding momentum correlations $k_{p_1\text{-HI}}-k_{p_2\text{-HI}}$, where HI corresponds to the $A-2$ nucleus. In the first case (a) of prompt $2p$ decay, sequential $2p$ emission should be energetically forbidden. As both emitted protons should share $2p$-decay energy $Q_{2p}$, their energy spectra are broad and centered around the value of $Q_{2p}/2$; consequently, the $2p$ momentum-correlation plot should have the shape of an arc, with a radius corresponding to the $Q_{2p}$ value and with most of the counts lying in the peak indicated by the dark spot in the lower panel of Fig.~\ref{fig:kinem_a_b_c} (a). Note that all momentum-correlation plots in Fig.~\ref{fig:kinem_a_b_c} are symmetric with the respect to the $45^\circ$ line since the protons $p_1$ and $p_2$ are indistinguishable.

The case (b) represents the sequential emission  of two protons through a narrow resonance in the intermediate nucleus with $A-1$. The proton energies are fixed here, and the $k_{p_1\text{-HI}}-k_{p_2\text{-HI}}$ correlation plot should yield double peaks as indicated by the black dots in the lower panel of Fig.~\ref{fig:kinem_a_b_c} (b).

The third $2p$-decay mechanism is 2\emph{p} emission from several broad continuum parent states via a low-lying state in $A-1$, see Fig.~\ref{fig:kinem_a_b_c} (c). This mechanism should reveal a peak in the $p$-HI energy with the corresponding broad distribution along the narrow ``slice'' as shown in the lower part of Fig.~\ref{fig:kinem_a_b_c} (c).

\begin{figure}[tb]
\begin{center}
\includegraphics[width=0.8\linewidth,angle=0.]{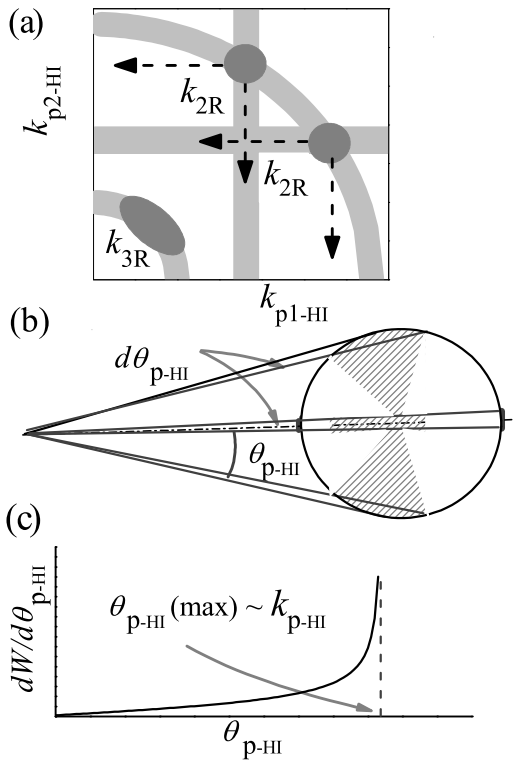}
\end{center}
\caption{(a) Cartoons of transverse momentum correlations $k_{p_1\text{-HI}}-k_{p_2\text{-HI}}$ for a case of two precursor states with prompt and sequential $2p$-decay mechanisms. (b) Kinematics of isotropic mono-energetic proton emission from a high-energy heavy ion HI. (c) The corresponding angular $p$-HI distribution exhibiting the peak, corresponding to the $Q$-value of the $1p$ decay.}
\label{fig:kinematics}
\end{figure}

In the present method, we measure only total HI momentum and relative $p$-HI angles in the transverse direction. We register trajectories of all decay products directly downstream from the secondary-reaction target. Fig.~\ref{fig:kinematics} (b) shows the kinematics plot for the simple case of isotropic and mono-energetic single-proton emission from a high-energy heavy ion. Fig.~\ref{fig:kinematics} (c) shows the corresponding distribution of laboratory $p$-HI opening angles, $\theta_{p\text{-HI}}$. The angular spectrum exhibits a sharp peak corresponding to the proton emitted almost orthogonal to the HI momentum vector. Thus the maximum value of $\theta_{p\text{-HI}}$ is directly related to the decay energy $Q_p$ of the emitted proton. In the same way, the $k_{p_1\text{-HI}} - k_{p_2\text{-HI}}$ momentum correlations for $2p$ decays (Fig.~\ref{fig:kinematics}) can be replaced by the corresponding $\theta_{p_1\text{-HI}} - \theta_{p_2\text{-HI}}$ correlations. If the initial and final states of $p$ emission are narrow, the width of a peak in the angular distribution is governed mostly by the angular straggling of the proton in the secondary-reaction target. If those states are broad, the width results from a convolution of the state's width with the proton angular straggling.

The cases, sketched in Figs.~\ref{fig:kinem_a_b_c} and \ref{fig:kinematics} represent ideal cases. In reality, several proton branches may be present, representing more than one of the cases, shown schematically in Fig.\ \ref{fig:kinematics} (a), and leading to a complicated spectra with several peaks. One can, however, clean up the spectra and enhance e.g.\ transitions with the small $Q_{2p}$-values by gating on the small angles of $\theta_{p_1\text{-HI}}$ and plotting the spectrum of $\theta_{p_2\text{-HI}}$ under this condition.

Another tool of data analysis is a kinematic variable
\[
\rho_\theta = \sqrt{\theta^2_{p_1\text{-HI}} + \theta^2_{p_2\text{-HI}}} \,,
\]
describing 3-body HI+$p$+$p$ angular correlations. Since $\rho_\theta$ is related to the energy sum of both emitted protons and, therefore, to the $Q_{2p}$ of the parent state by the relation $Q_{2p} \sim \rho^2_\theta$ \cite{Mukha:2012}, one can obtain an indication of the parent state and its $2p$-decay energy by studying the distribution of $\rho_\theta$. In a case of the decay from the same state, two protons share $Q_{2p}$, and such events should be located along a root-mean-square arc in an angular correlation plot $\theta_{p_1\text{-HI}}-\theta_{p_2\text{-HI}}$. By gating on a particular $\rho_\theta$ arc, the decay events from a certain $2p$-precursor can be selected. The $\rho_\theta$ distributions are very useful in the analysis of $2p$-decay data since they produce the spectra with less peaks and allow to gate on a specific excitation-energy regions.

In all cases, detailed Monte-Carlo simulations are required in order to interpret the angular spectra quantitatively by taking into account the corresponding response of the experimental setup. For example, the angular correlation for fixed energy decay must be first calculated. This predicted angular correlation is then compared to the measured one. The resonance energy is obtained by the best-fit where the probability that the two distributions are identical has maximum value. In the same way, limits for the width of a resonance can be obtained.

The above-described analysis procedure, where the states observed in a $2p$-precursor were investigated by comparing the measured angular $\theta$ and $\rho_{\theta}$ correlations with the Monte Carlo (MC) simulations of the respective detector response, has been published in Refs.\ \cite{Mukha:2010,Mukha:2012}. We follow this procedure in the present work, and the applied detector calibrations are taken from the previous $^{30}$Ar analysis of the same experiment \cite{Xu:2018}.


\subsection{The data analysis: unknown states in $^{29}$Ar and $^{28}$Cl}


\begin{figure}[t!]
\begin{center}
\includegraphics[width=0.45\textwidth]{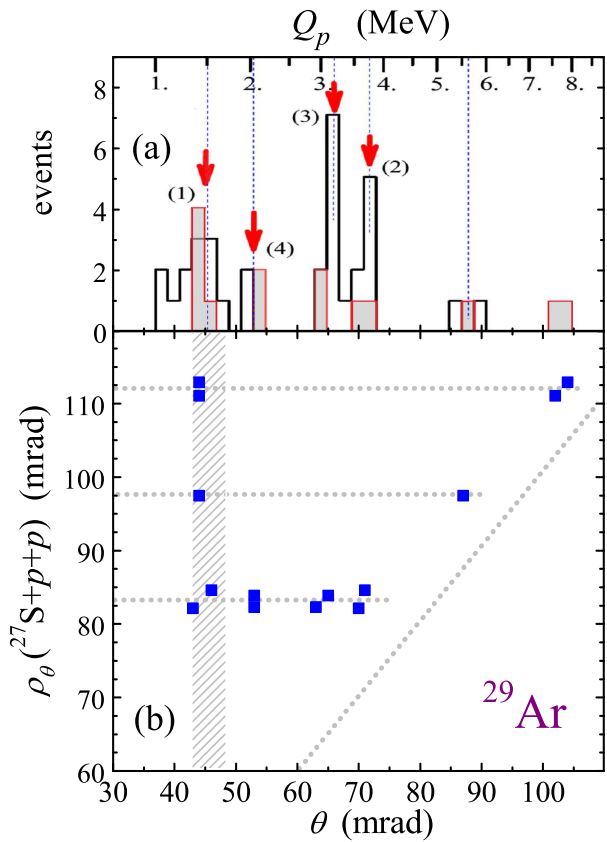}
\end{center}
\caption{  The angular correlation plot $\theta - \rho_{\theta}$ (angles are in mradians) for $^{29}$Ar decay via $^{27}$S+$p$+$p$ channel is shown in (b). The hatched area highlights events from the decay of a $^{27}$S+$p$ resonance assigned as the $^{28}$Cl ground state. The dotted lines guide the eye for the regions corresponding to assumed $^{29}$Ar states. (a) shows the angular correlations $\theta_{p\text{-}^{27}\text{S}}$ (shaded histogram) obtained as projection on the $\theta$ axis from the $^{27}$S+$p$+$p$ channel of (b). The $\theta_{p\text{-}^{27}\text{S}}$ ``inclusive'' angular correlations obtained from the measured $p$-$^{27}$S coincident events are shown by the black hollow histogram. The labeled (1)--(4) arrows highlight the events inspected for the possible $^{28}$Cl resonance states. The upper axis shows the corresponding $1p$-decay energies $Q_p$. The 45$^\circ$-tilted dotted line in (b) shows the kinematical limit.}
\label{fig:exp-29ar}
\end{figure}

We begin from the analysis of the relatively simple case of the measured $^{27}$S+$p$+$p$ correlations presented by the $\theta_{p_1\text{-}^{27}\text{S}}-\theta_{p_2\text{-}^{27}\text{S}}$ and $\theta-\rho_{\theta}$ plots in Figs.~\ref{fig:Ang1_Ang2_S27pp} and \ref{fig:exp-29ar}(b), respectively. These $^{29}$Ar-related correlations   comprise just seven 2\emph{p}-decay events, each being measured in triple $^{27}$S\emph{+p+p} coincidence. Each detected event provides two $\theta_{p\text{-}^{27}\text{S}}$  and one $\rho_\theta$ values. Most of them are very well focused around the locations at $\theta=44$~mrad or $\rho_{\theta}=84$~mrad. These values correspond to the 1\emph{p}-decay of the $^{28}$Cl state with $E_r$ of about 1.6 MeV and to the 2\emph{p}-decay of the $^{29}$Ar state with $Q_{2p}$ of about 5.5 MeV. A cross-check of this conclusion is illustrated in Fig.\ \ref{fig:exp-29ar}(a) where the angular correlations $\theta_{p\text{-}^{27}\text{S}}$ projected from the $^{27}$S+$p$+$p$ correlation plot [in Fig.\ \ref{fig:exp-29ar}(b)]  are compared with the ``inclusive'' $\theta_{p\text{-}^{27}\text{S}}$ distribution obtained from the measured  $p\text{-}^{27}\text{S}$ double-coincidence events. One may see that the ``inclusive'' spectrum consists of relatively enhanced peaks (1--3). The peaks (1) and (2) have the best-fits at the 1\emph{p}-decay energies $E_r$ of 1.60(8) and 3.9(1) MeV, respectively. They have been assigned as the first- and second-emitted protons from the 5.5 MeV state in $^{29}$Ar, and their sum decay energy gives the total $2p$-decay energy of 5.50(18) MeV.

\begin{figure}
\begin{center}
\includegraphics[width=0.48\textwidth]{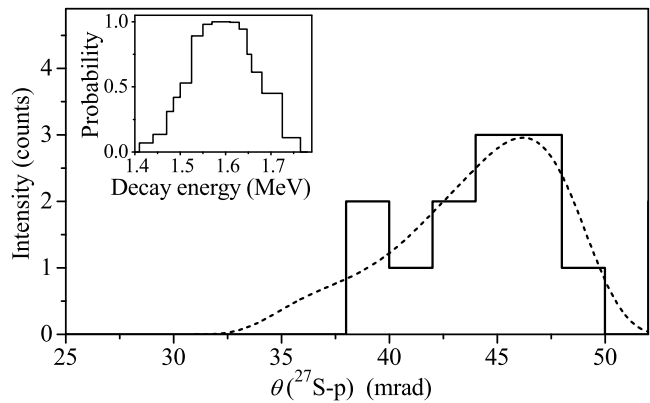}
\end{center}
\caption{Best fit of the peak (1) of the $\theta_{p\text{-}^{27}\text{S}}$ ``inclusive'' angular correlations from Fig.~\ref{fig:exp-29ar}(b) (histogram) by simulations of the setup response to in-flight decays of $^{28}$Cl with the 1\emph{p}-decay energy of 1.60 MeV (dashed curve). The inset shows probability that the simulated distribution matches the data as function of the 1\emph{p}-decay energy.}
\label{fig:S27-p_fit}
\end{figure}

The data-fitting procedure is illustrated on the example of the (1) peak at $\theta=44$~mrad in the $p\text{-}^{27}\text{S}$ correlation in Fig.~\ref{fig:S27-p_fit}. This is the same procedure described in details in Refs.~\cite{Mukha:2010,Xu:2018}. The best-fit simulations obtained for the in-flight decay of $^{28}$Cl with the 1\emph{p}-decay energy of 1.60 MeV describe the data quantitatively, and the figure inset shows that the probability of the data matching simulations is about 1. The full width at half maximum of the probability distribution provides the evaluation of the $E_r$ uncertainty.

 There are two additional events in the decay patten of the 5.5 MeV state in Fig.\ \ref{fig:exp-29ar} (b) corresponding to the inclusive peaks (3) and (4) in Fig.\ \ref{fig:exp-29ar} (a). As the inclusive peak (3) is much enhanced, we may speculate that it may be an evidence on the second state in $^{28}$Cl, which is also fed by the other unspecified reaction channels, illustrated in Fig.\ \ref{fig:kinem_a_b_c}(c). The best-fit 1\emph{p}-decay energy of the peak (3) is 3.20(6) MeV.

\begin{figure}
\begin{center}
\includegraphics[width=0.45\textwidth]{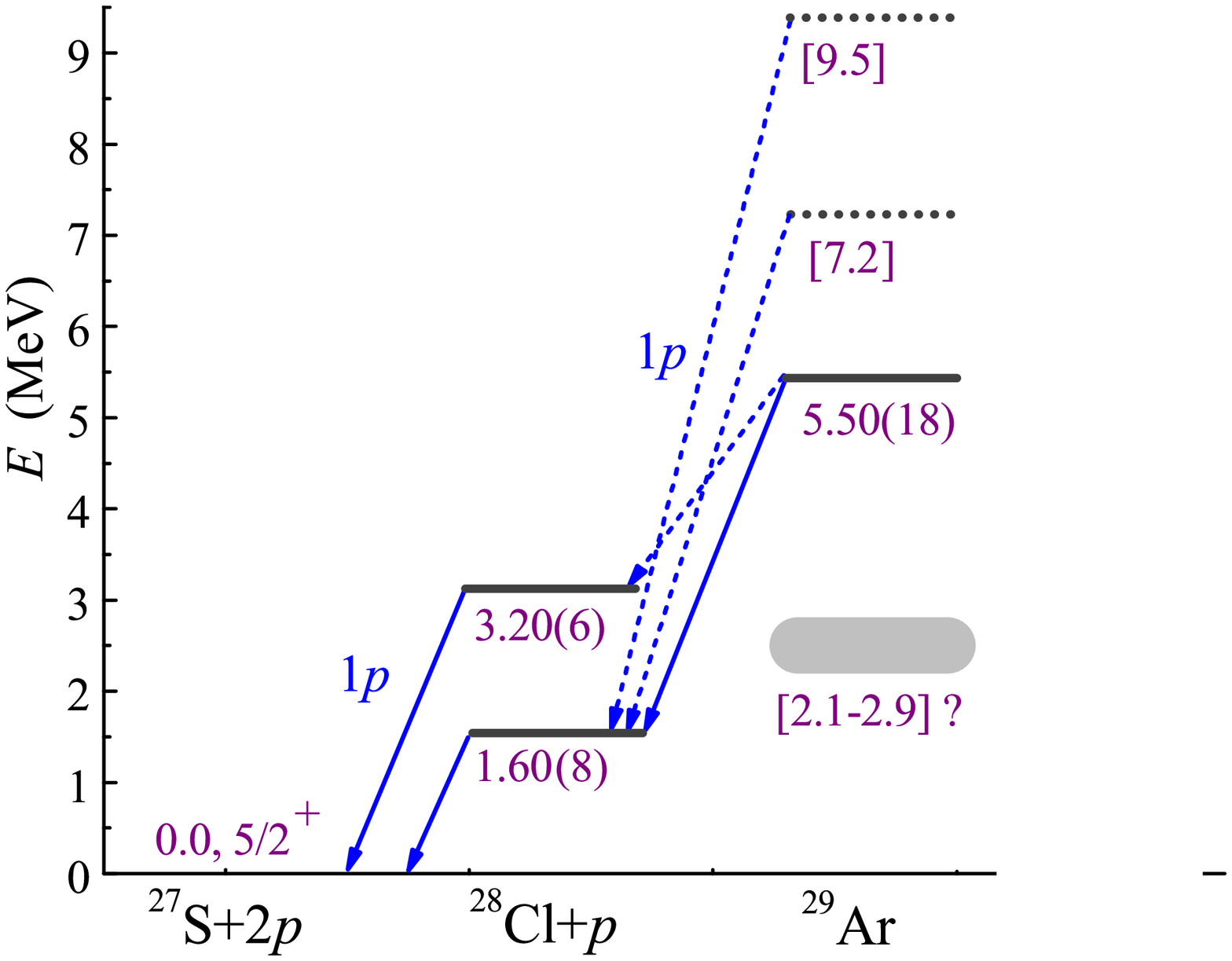}
\end{center}
\caption{  The suggested decay scheme and the energy levels of $^{29}$Ar and $^{28}$Cl relative to the $^{27}$S+2\emph{p} threshold. The gray-shaded region corresponds to the ground state energy of $^{29}$Ar predicted in this work, see the discussion in section \ref{sec:sep-cl-ar}. The assumed and indicated $1p$ transitions are shown by the solid and dotted arrows, respectively. The dotted levels indicate very tentative assignments made on the basis of one or two $2p$-decay events.}
\label{fig:decay_scheme_29ar}
\end{figure}

In addition, there are indications on $^{27}$S+$p$+$p$ correlations at $\rho_{\theta}$ of 97 and 112~mrad, which may correspond to the  $2p$-decays of $^{29}$Ar with $Q_{2p}$ of about 7.2 and 9.5~MeV, respectively. Both of the indicated states have the second-emitted proton energy of 1.6 MeV, which corresponds to the lowest assigned state in $^{28}$Cl.

The derived decay scheme and levels of $^{29}$Ar and $^{28}$Cl are shown in Fig.\ \ref{fig:decay_scheme_29ar}.

\begin{figure}[t!]
\begin{center}
\includegraphics[width=0.45\textwidth]{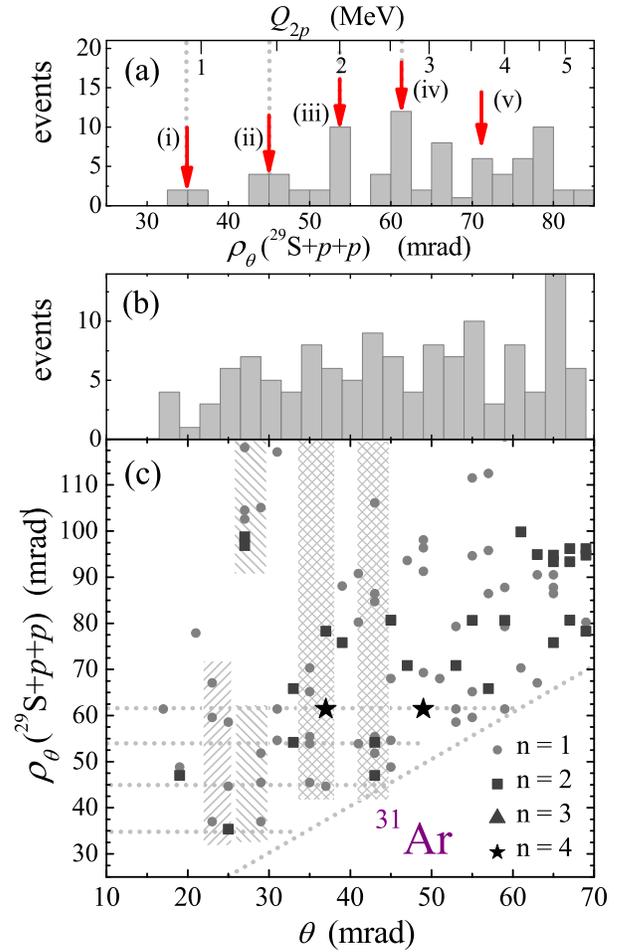}
\end{center}
\caption{  The angular correlation plot $\theta$-$\rho_{\theta}$ derived for the measured $^{29}$S+$p$+$p$ channel (c). The (a) and (b) show  corresponding projections on the $\rho_{\theta}$ and $\theta$ axes, respectively. The hatched areas in (c) highlight events assumed to originate from the decay of $^{30}$Cl resonance states. The dotted lines in the (a) and (c) guide the eye for the regions corresponding to the assumed $^{31}$Ar states which correspond to the events highlighted by the arcs in Fig.~\ref{fig:Ang1_Ang2_S29pp} labeled by the Roman numerals. The arrows and vertical lines in (a) point to the regions inspected for possible states in  $^{31}$Ar. The cases (i)-(v) correspond to the events highlighted by the arcs (i)-(v) in Fig.~\ref{fig:Ang1_Ang2_S29pp}, respectively. The upper axis shows the $2p$ decay energy $Q_{2p}$ in the $^{29}$S+$p$+$p$ system.  The inclined dotted line in (c) shows the kinematic limit for this type of plot; shape coding corresponds to multiplicity of events.}
\label{fig:exp-31ar}
\end{figure}

We argue below in Section III, that our empirical assignments are backed by the isobaric mirror symmetry systematics and that the most probable interpretation of the measured decay-product correlations is the observation of $^{28}$Cl ground state with $S_p$ = -1.60(8)~MeV and the $^{29}$Ar excited state with $S_{2p}$ = -5.50(18)~MeV.


\subsection{The data analysis: unknown states in $^{31}$Ar and $^{30}$Cl}


The $2p$-decay pattern of $^{31}$Ar, derived from the $^{29}$S+$p$+$p$ data, is more complicated. Several separated regions with events, corresponding to the same $2p$-decay energy, can be distinguished at the low angles in Fig.~\ref{fig:Ang1_Ang2_S29pp}, which indicate different states in $^{31}$Ar. The tentatively selected arcs are labeled by the Roman numerals (i)--(iv). The same event groups can be found in the angular $\theta-\rho_{\theta}$ correlation plot in Fig.~\ref{fig:exp-31ar} (c) derived for the assumed $^{31}$Ar $2p$-decays. Its projections on the $\theta$ and $\rho_{\theta}$ axes are shown in the panels (b) and (a) in Fig.~\ref{fig:exp-31ar}, respectively.
 The $\theta$($^{29}$S-$p$) projection indicates some structures centered at the angles $\theta =\{26,37,43\}$~mrad, which point to possible low-energy states in $^{30}$Cl. The $\rho_\theta$ projection indicates several  $2p$-decay patterns in $^{31}$Ar with the centre-of-gravity values at $\rho_{\theta}=\{45,53, 61\}$~mrad.

\begin{figure}[t!]
\begin{center}
\includegraphics[width=0.47\textwidth]{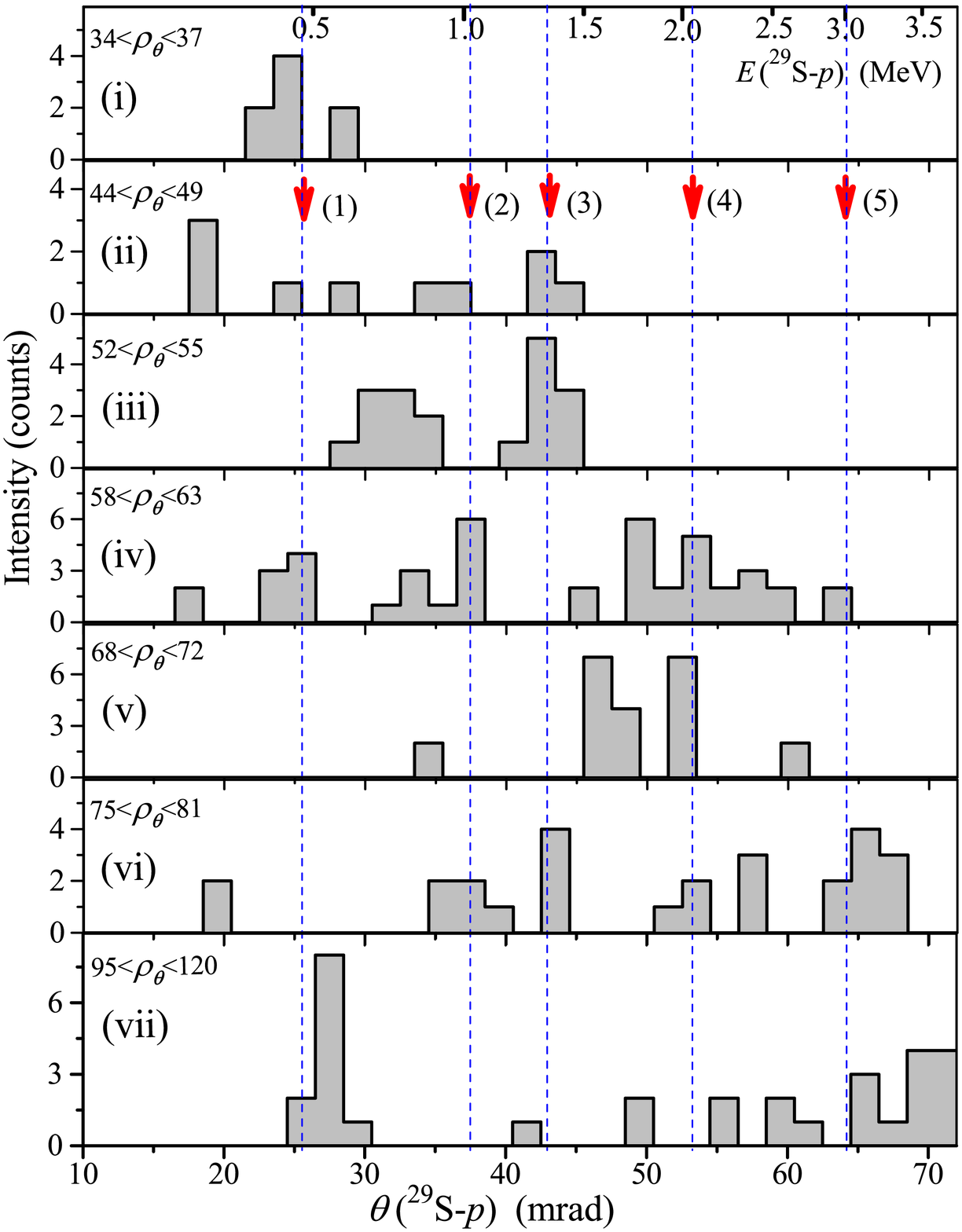}
\includegraphics[width=0.47\textwidth]{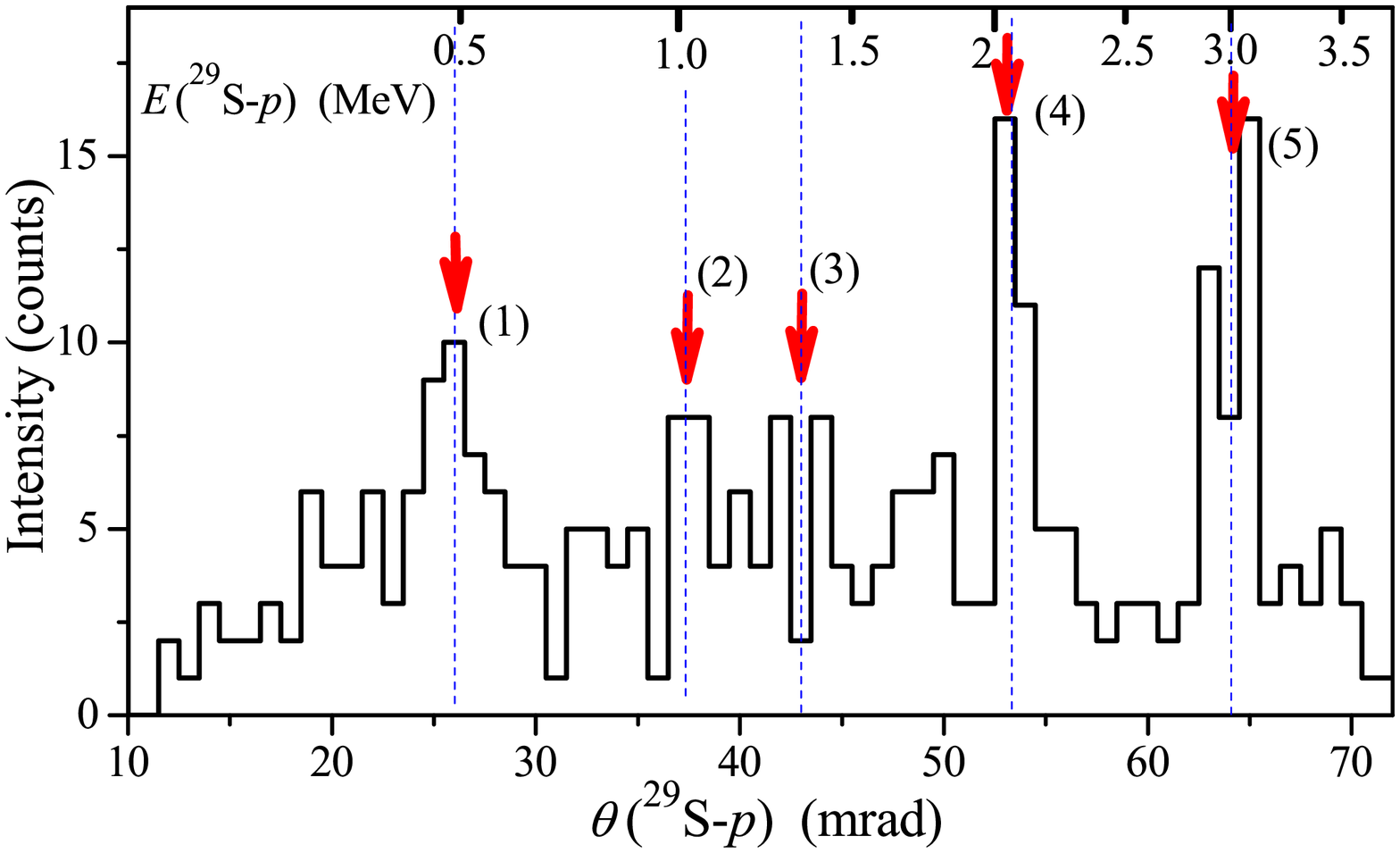}
\end{center}
\caption{  The ``gated'' angular correlations  $\theta_{p\text{-}^{29}\text{S}}$ derived from the measured $^{29}$S+$p$+$p$ triple-coincidence events, which are selected  by choosing the gate conditions within the $\rho_\theta$ ranges corresponding to the highlighted  arcs in Fig.~\ref{fig:Ang1_Ang2_S29pp}. The panels (i)--(iv) correspond to the selection gates labeled in Fig.~\ref{fig:Ang1_Ang2_S29pp}  by the same Roman numerals. The additional panels (v)--(vii) present the similar $\theta_{p\text{-}^{29}\text{S}}$ correlations selected by  the larger  $\rho_\theta$ values shown in their upper-left corners.  The panel (viii) shows the ``inclusive'' angular $\theta_{p\text{-}^{29}\text{S}}$  correlations obtained from the measured  $^{29}$S+$p$ double coincident events (the hollow histogram). The upper axes show the corresponding  energies in the $^{29}$S+$p$ system. The arrows (1)--(5) point to the events inspected for possible  resonances in $^{30}$Cl as well as the vertical  across-panel lines.}
\label{fig:29Spp_theta_gated}
\end{figure}

 The obtained statistics  of the measured triple coincidences is low, and the non-selective projections do not allow for a quantitative analysis. Thus we have used the slice $\theta$ projections gated by the $\rho_\theta$ selected areas (i-v) in Fig.~\ref{fig:exp-31ar} (a).
 These gated projections are shown in Fig.~\ref{fig:29Spp_theta_gated} in the panels (i-v), respectively. Two additional projections gated at very large $\rho$ values are shown in the panels (vi,vii).
 In analogy to the $^{29}$Ar analysis, the ``inclusive'' $\theta_{p\text{-}^{29}\text{S}}$ distribution obtained from the measured  $p\text{-}^{29}\text{S}$ double-coincidence events is shown in the lowest panel of Fig.~\ref{fig:29Spp_theta_gated}. This inclusive distribution display the same low-energy peak (1) at around 26~mrad as well as the peaks (4,5), though evidence on the Fig.~~\ref{fig:exp-31ar}(c)-indicated peaks  at 37 and 43~mrad (marked as (2) and (3), respectively) is weak.

Similarly to the $^{28}$Cl case, the MC simulations of the well-distinguished peaks (1,4,5) in the lowest panel of Fig.~~\ref{fig:29Spp_theta_gated}  by the experimental-setup response have resulted in assigning of the unknown $^{30}$Cl states with the $1p$-decay energies $E_r$ of 0.48(2), 2.00(5) and 3.0(2) MeV, respectively. On the basis of the performed analysis,
the  0.48(2)~MeV peak is assumed to be the ground state of $^{30}$Cl. Such an assignment is supported by the observed $^{29}$S+$p$+$p$ correlations where one of the emitted protons has relatively large energy and another proton's energy is peaked at around 0.5 MeV, see Fig.~~\ref{fig:29Spp_theta_gated}(vii). This is a typical  situation for a final-state interaction due to the $^{30}$Cl g.s.\ resonance, see illustration in Fig.~\ref{fig:kinem_a_b_c}(c).

\begin{figure}
\begin{center}
\includegraphics[width=0.48\textwidth]{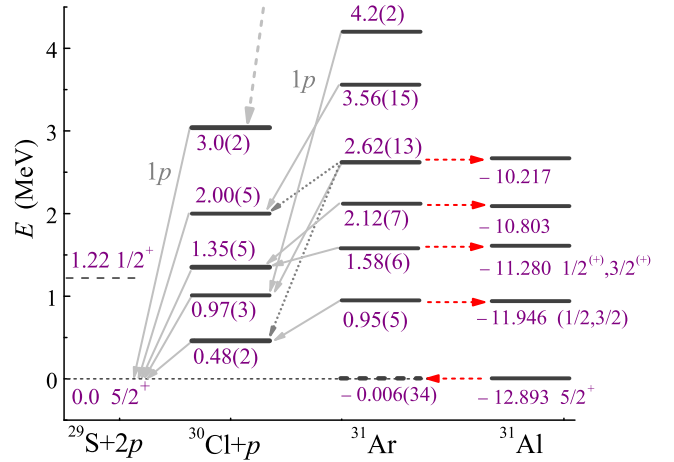}
\end{center}
\caption{  The decay and level schemes for $^{31}$Ar and $^{30}$Cl isotopes derived from the data. The assigned $1p$ transitions are shown by the light gray solid arrows. The dotted arrows show two undistinguished decay branches of the 2.62(13) state in $^{31}$Ar, and the dashed arrow indicates 1\emph{p} transitions from unidentified states in $^{31}$Ar feeding the 3.0(2) state in $^{30}$Cl. Vertical axis shows the energies relative to proton (for $^{30}$Cl) and two-proton (for $^{31}$Ar) breakup thresholds. The four lowest excited states of isobaric mirror partner $^{31}$Al are aligned with corresponding observed states of $^{31}$Ar (the correspondence of the levels is shown by red dashed arrows) and the $^{31}$Ar g.s.\ energy is inferred based on isobaric symmetry assumptions. The legends for $^{31}$Al levels show energies relative to the $2p$-breakup threshold and spin-parity $J^{\pi}$ of the state.}
\label{fig:decay_scheme_31ar}
\end{figure}

By using the parameters of the $^{30}$Cl g.s.\, one may obtain the 2$p$-decay energy of the lowest-energy state in $^{31}$Ar observed in the $^{29}$S+$p$+$p$ correlations, see Fig.~\ref{fig:29Spp_theta_gated}(i).
We have fitted the $\theta$ projection (i) by a sequential proton decay of  $^{31}$Ar via the g.s.\  of $^{30}$Cl, and the obtained value of 2$p$-decay energy is 0.95(5)~MeV.
 Though the pattern centered at $\rho_{\theta}=35$~mrad has low statistics, it is very important for an overall interpretation of the data, as it highly likely corresponds to the $^{31}$Ar first excited state.
Thus we may lay the first piece into the puzzle of the $^{31}$Ar excitation spectrum and its $2p$-decay scheme whose complete reconstruction is shown in Fig.~\ref{fig:decay_scheme_31ar} and  which is explained in a step-by-step way below.
\begin{itemize}
  \item{Namely, the  gated  $\theta$ projections in  Fig.~\ref{fig:29Spp_theta_gated} (ii) and (iii) exhibit the same  peak (3) at 43~mrad, which point to the sequential 2\emph{p} decays of these $^{31}$Ar states via the same state in $^{30}$Cl.
    The peak (3) is best-fitted by assuming the 1\emph{p} decay of the $^{30}$Cl state with $E_r$=1.35(5)~MeV. Then the  $^{31}$Ar states corresponding to the complementary bumps in the structures (ii) and (iii) have the fitted 2\emph{p}-decay energies of 1.58(6) and 2.12(7)~MeV, respectively. One should note that the projection (ii) provides very broad and statistically poor signal from the corresponding $^{31}$Ar state, which makes the  assignment  very tentative, see Fig.~\ref{fig:decay_scheme_31ar}.}
  \item{Next, the  gated  $\theta$ projections in  Fig.~\ref{fig:29Spp_theta_gated} (iv) and (v) reveal events matching the same 2.00~MeV peak (4) in the inclusive spectrum in the lowest panel in Fig.~\ref{fig:29Spp_theta_gated}. They point to the sequential 2\emph{p} decays  of two more  states in $^{31}$Ar via the  2.00~MeV state in $^{30}$Cl. In particular, the fit of the peak  at $\sim$48~mrad which is complementary to the peak (4) in the projection (v) yields its energy of 1.56(10)~MeV, and together they allow for assignment of the new $^{31}$Ar state with the 2\emph{p}-decay energy of 3.56(15)~MeV, see Fig.~\ref{fig:decay_scheme_31ar}. Interpretation of the  $\theta$ projections in  Fig.~\ref{fig:29Spp_theta_gated} (iv) is more complicated, because it has the additional components, and one of them matches the peak (2) at 37~mrad due to a  suspected state  in $^{30}$Cl.}
  \item{The contribution of such a state can be spotted also in the $\theta$ projection (vi) in  Fig.~\ref{fig:29Spp_theta_gated} as well as in the ``inclusive''  $\theta$ distribution labeled as (2). The corresponding fits provide the 1\emph{p}-decay energy of 0.97(3)~MeV assigned to the $^{30}$Cl state. Then the whole structure of the $\theta$  distribution (iv) in Fig.~\ref{fig:29Spp_theta_gated} may be explained by a sequential 2\emph{p}-decay   of one state in $^{31}$Ar by two  branches via the intermediate 0.97 and 2.00 states in $^{30}$Cl. The respective fits provide two independent evaluations of the 2\emph{p}-decay energy of  the $^{31}$Ar  state of 0.97(3)+1.65(10)=2.62(13) and 2.00(5)+0.45(3)=2.45(8) MeV, respectively. They agree within the statistical uncertainties. One may note that the assigned 2\emph{p}-decay branch via the 2.00~MeV state in $^{30}$Cl has the first-emitted proton energy of 0.45(3)~MeV, which coincides with the 1\emph{p}-decay energy of the g.s.\  of $^{30}$Cl. Therefore the sequential 2\emph{p} decay may proceed also via the g.s.\  of $^{30}$Cl. These two assignments indistinguishable in our experiment are shown in Fig.~~\ref{fig:decay_scheme_31ar} by the dotted arrows.  Due to this uncertainty, we accept the $^{31}$Ar  state to be at 2.62(13)~MeV.}
  \item{Finally, the  gated  $\theta$ projection in  Fig.~\ref{fig:29Spp_theta_gated} (vi) with the assumed   peak (2) due to the 0.97~MeV state in $^{30}$Cl allows for identification of the highest state observed in $^{31}$Ar with the 2\emph{p}-decay energy of 0.97(3)+3.2(2)=4.2(2)~MeV.}
  \item{The only undiscussed peak (5) at about 65~mrad in the ``inclusive''  $\theta$ distribution in the lowest panel of Fig.~\ref{fig:29Spp_theta_gated} is also detected in the observed $^{29}$S-$p$-$p$ correlations, see Fig.~\ref{fig:exp-31ar}(b). However, energy of another emitted proton is distributed in a broad range of energy, which points to a  continuum region of $^{31}$Ar excitations above 5~MeV. Therefore the peak (5) can not be assigned to an individual $^{31}$Ar state. We may speculate that it probably belongs to the 3.0(2)~MeV state in $^{30}$Cl which is strongly populated by de-excitation of high-energy continuum in $^{31}$Ar.}
\end{itemize}

Summarizing the above considerations, we have assigned the  $^{30}$Cl states with the decay energies $E_r$ of 0.48(2), 0.97(3), 1.35(5), 2.00(5) and 3.0(2)~MeV. There is also some indication that the structure around $\theta =26$~mrad  may consist of two sub-structures at about 24 and 28 mrad (corresponding to the $E_r$ values of 0.48 and 0.55 MeV, respectively), which we will discuss below.
The newly prescribed states in $^{31}$Ar have the 2\emph{p}-decay energies of 0.95(5), 1.58(6), 2.12(7), 2.62(13), 3.56(15) and 4.2(2)~MeV.
All derived  levels in $^{31}$Ar and $^{30}$Cl and their decay transitions are shown in Fig.\ \ref{fig:decay_scheme_31ar}.

\section{Systematics for chlorine isotopes}
%

\begin{figure}
\begin{center}
\includegraphics[width=0.48\textwidth]{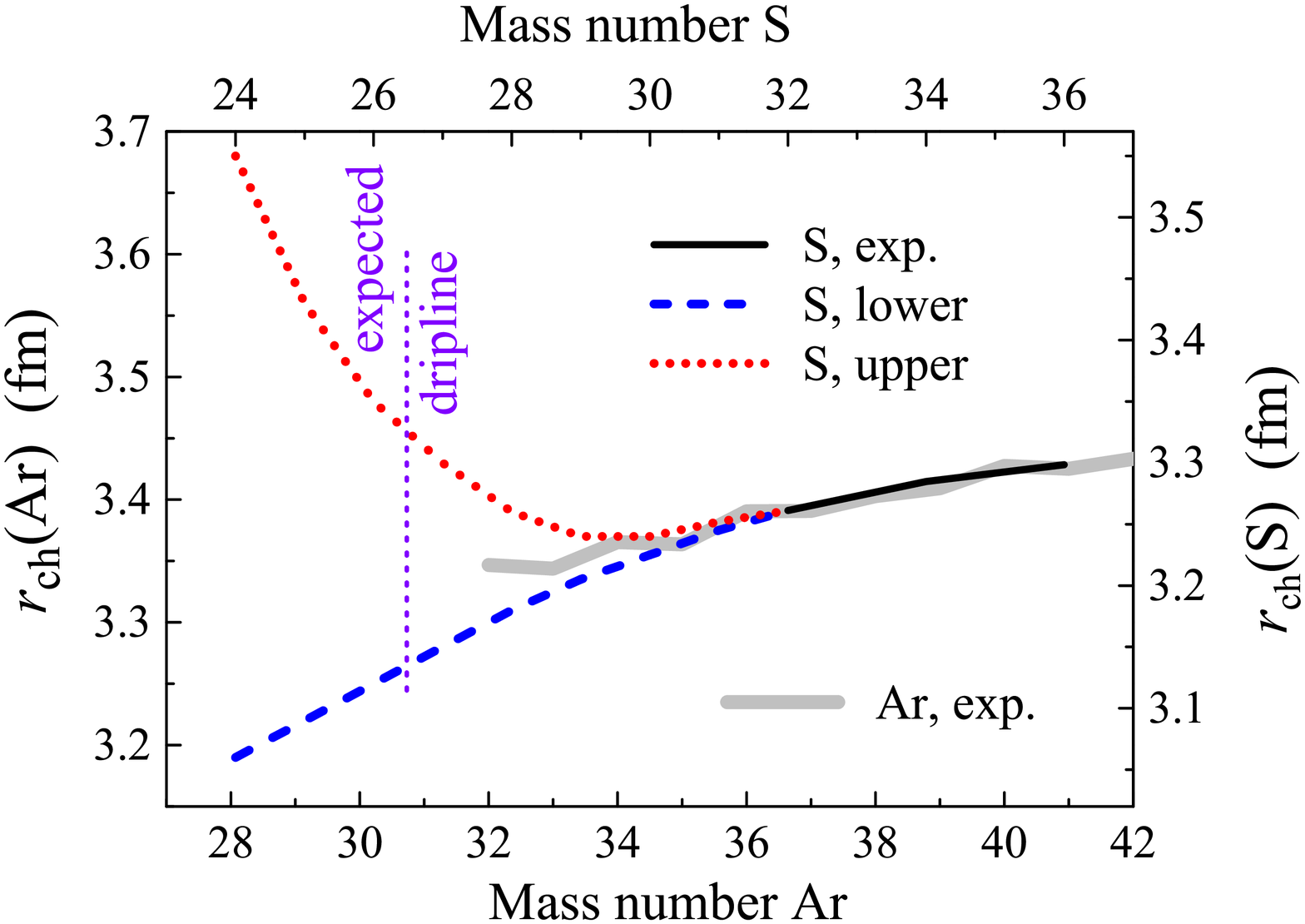}
\end{center}
\caption{  Charge radii of sulphur isotopes used in the cluster core+$p$ model for chlorine states. The dependence for sulphur isotopes is aligned with the much better studied dependence for argon isotopes to substantiate the provided extrapolation.}
\label{fig:s-charge-radii}
\end{figure}

As a first step in the interpretation of the data, we would like to evaluate the energies of the states in proton-rich chlorine isotopes systematically by using the known information about their isobaric mirror partners. The obstacle here is the Thomas-Ehrman shift (TES) effect \cite{Ehrman:1951,Thomas:1952}, especially pronounced in the $s$-$d$ shell nuclei. The systematics of orbital size variations for $s$- and $d$-wave configurations are different when approaching the proton dripline and beyond it. This leads to a significant relative shift of the $s$-wave and $d$-wave dominated states distorting the expected (due to isobaric symmetry) nuclear level ordering in isotopes near the proton dripline. The core+$p$ cluster model is a reasonable tool for consideration of this effect.

The Coulomb displacement energies in the core+$p$ cluster model depend on two parameters: the orbital radius, which is mainly controlled by the potential radius, and the charge radius of the core. We use the potential with a Woods-Saxon formfactor and with a conventional diffuseness parameter $a=0.65$ fm. The radius value is provided by the standard parameterizations
\begin{equation}
r_0=1.2 (A_{\text{core}}+1)^{1/3}\,.
\label{eq:pot-rad}
\end{equation}
The charge radii of sulphur isotopes are poorly studied \cite{Angeli:2013}, so we use the extrapolation shown in Fig.\ \ref{fig:s-charge-radii}. Here we use two limits, corresponding to either ascending or descending trend near the dripline (both trends are not excluded by the available systematics of the charge radii). One should note that the $^{26}$S case is already uncertain. This particle-unstable nuclide (expected to be a $2p$-precursor \cite{Fomichev:2011}) has the valence-proton wave function expected to well penetrate into the sub-barrier region.

Then the Coulomb potential of the charged sphere is used with the radius parameter  $r_{\text{sph}}$,
\begin{equation}
r^2_{\text{sph}}=(5/3)[r^2_{\text{ch}}(A_{\text{core}}) + r^2_{\text{ch}}(p)]\,,
\label{eq:sph-rad}
\end{equation}
where $r_{\text{ch}}(p) = 0.8$ fm. The potential parameters are collected in Table \ref{tab:poten-param}. The results of the calculations are collected in Fig.\ \ref{fig:levels-cl}. Below, we study the chain  of five chlorine isotopes $^{32-28}$Cl.

\begin{table}[b]
\caption{Potential parameters for the cluster two-body $^{A}$Cl=$^{A-1}$S+$p$ model. The  minimum and maximum $r_{\text{sph}}$ values correspond to the charge radii estimated from Fig.\ \ref{fig:s-charge-radii}.}
\begin{ruledtabular}
\begin{tabular}[c]{cccccccc}
$A$  & $r_0$ & $r_{\text{sph}}(\min)$ & $r_{\text{sph}}(\max)$  & $V_s$ (MeV) &  $V_d$ (MeV)  \\
\hline
32  & 3.81 & 4.31 & 4.32 &  & $-46.80$  \\
31  & 3.77 & 4.29 & 4.31 & $-45.00$ & $-45.38$ \\
30  & 3.73 & 4.26 & 4.31 & $-45.10$ & $-44.76$  \\
29  & 3.69 & 4.23 & 4.33 & $-41.87$ & $-41.85$  \\
28  & 3.64 & 4.20 & 4.38 & $-42.52$ & $-42.69$  \\
27  & 3.60 & 4.16 & 4.45 & $-34.85$ & $-39.51$  \\
26  & 3.56 & 4.12 & 4.55 & $-32.80$ & $-38.86$  \\
\end{tabular}
\end{ruledtabular}
\label{tab:poten-param}
\end{table}


\subsection{$^{31}$Cl and $^{32}$Cl cases}


One can see in Fig.\ \ref{fig:levels-cl} (a,b), that for known isotopes $^{31}$Cl and $^{32}$Cl the used systematics of potential parameters given by Eqs.\ (\ref{eq:pot-rad}) and (\ref{eq:sph-rad}) provides level energies which are   overbound a bit (by $\sim 150$ keV) in comparison with the data. However, the general trend is well reproduced, thus the standard set of the parameters could be the good starting point for the systematic evaluation of the whole isotope chain.


\subsection{$^{30}$Cl and $^{29}$Cl cases}


Spectrum of $^{29}$Cl was discussed in details in \cite{Mukha:2015,Xu:2018}, see Fig.\ \ref{fig:levels-cl} (d). The data on $^{30}$Cl spectrum is reported in this work for the first time. The spectra of these isotopes can be reasonably interpreted only on the bases of the strong TES effect for some states. The calculated levels shown in Fig.\ \ref{fig:levels-cl} (c) present evidence that two low-lying structures in the spectrum of $^{30}$Cl (at 0.48(2) and 0.97(3) MeV) can be associated with nearly-overlapping doublets $2^+$--$3^+$ and $1^+$--$3^+$. We assume that the $3^+$ g.s.\ in $^{30}$Al has a $d$-wave structure. Then its doublet partner, the $2^+$ state is expected to be strongly shifted down by TES, and therefore to  become the $^{30}$Cl g.s. There is a hint in the data shown in Figs.\ \ref{fig:exp-31ar} and \ref{fig:29Spp_theta_gated}, that the ``ground state peak'' in $^{30}$Cl at $\theta=26$ mrad actually consists of two substructures, differently populated in the decays of several $^{31}$Ar states. In this work, the $^{30}$Cl g.s.\ prescription is based on the lower substructure with the corresponding proton emission energy $E_r=0.48$ MeV.

Why the above-mentioned prescription is reliable? The Thomas-Ehrman shift for the $^{30}$Al-$^{30}$Cl g.s.\ pair is about 330 keV. If we assume that the $3^+$ g.s.\ in $^{30}$Al has an $s$-wave structure, then the Thomas-Ehrman shift leads to the evaluated energies $E_r=50-150$ keV of the $3^+$ g.s.\ in $^{30}$Cl. For such low decay energies, the $^{30}$Cl g.s.\ should live sufficiently long time in order to ``survive'' the flight through the second achromatic stage of the FRS fragment separator (of $\sim 150$ ns). We don't report such an experimental observation. We may also assume a $d$-wave structure of the $2^+$ and second $3^+$ states. However such an assumption practically does not change the predicted $S_p$ energy of $^{30}$Cl, but it requires the existence of peaks which are not seen in our data.


\subsection{$^{28}$Cl case}

\begin{figure*}
\begin{center}
\includegraphics[width=0.45\textwidth]{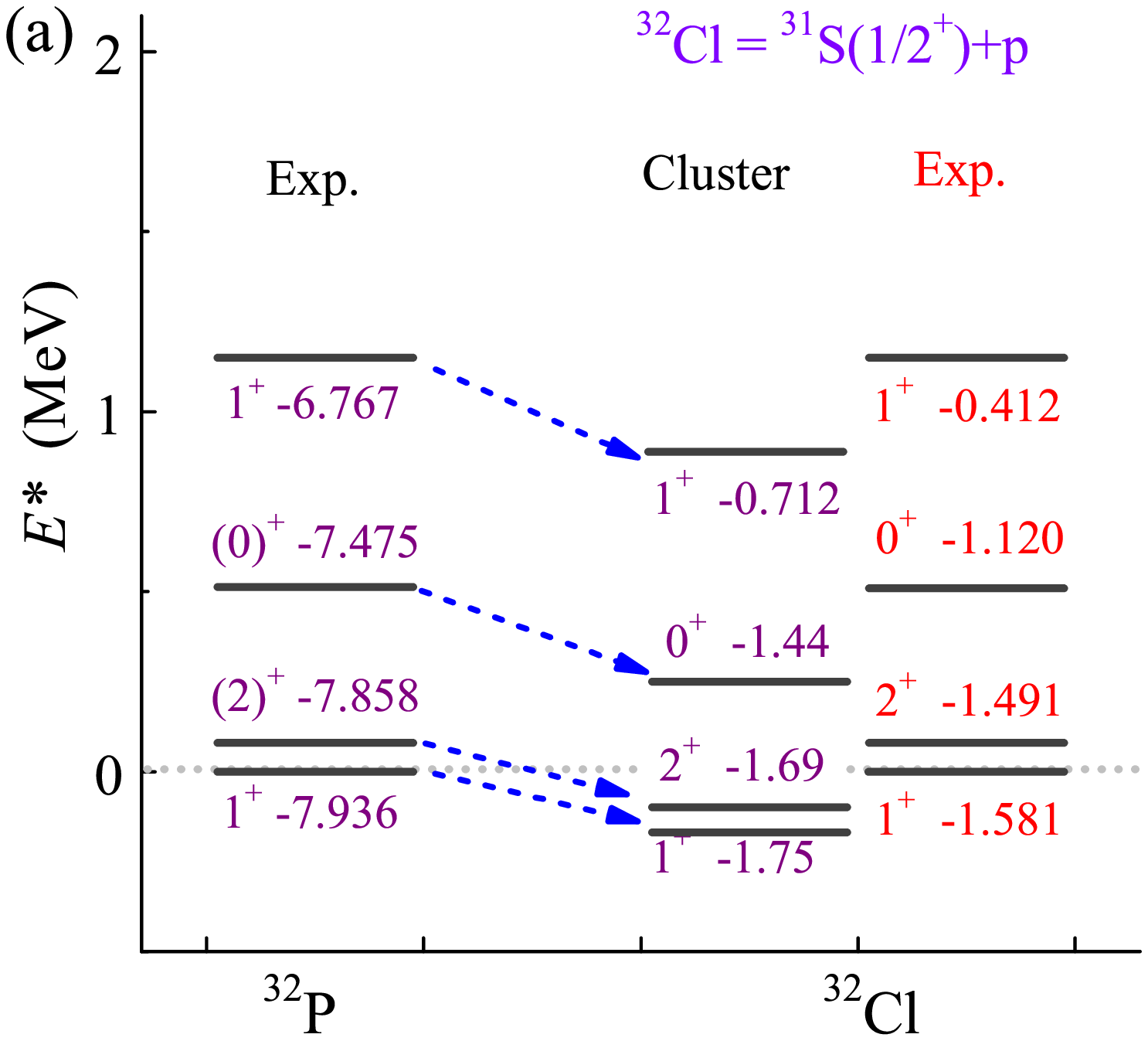} $\qquad$
\includegraphics[width=0.45\textwidth]{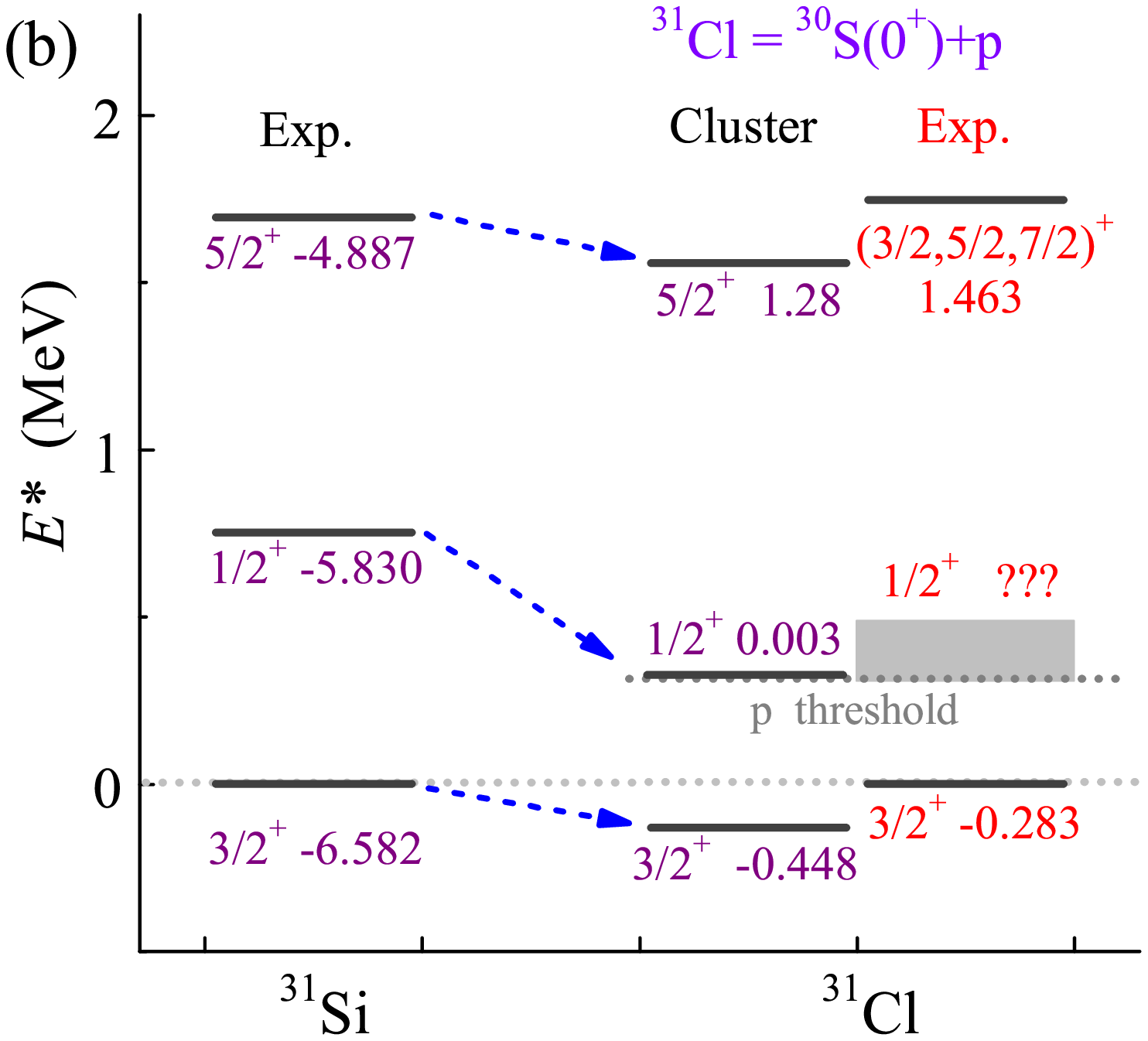}
\includegraphics[width=0.45\textwidth]{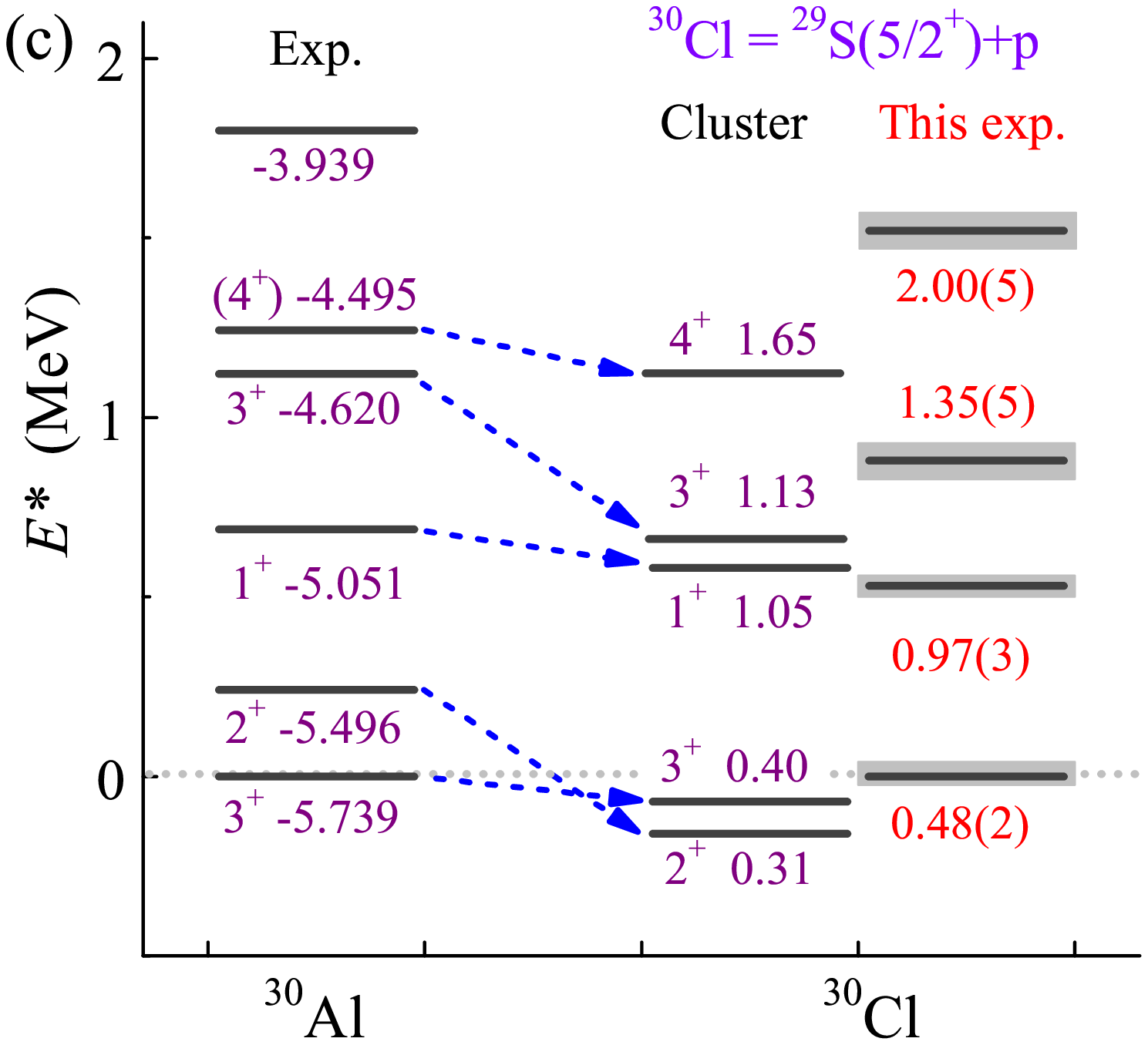} $\qquad$
\includegraphics[width=0.45\textwidth]{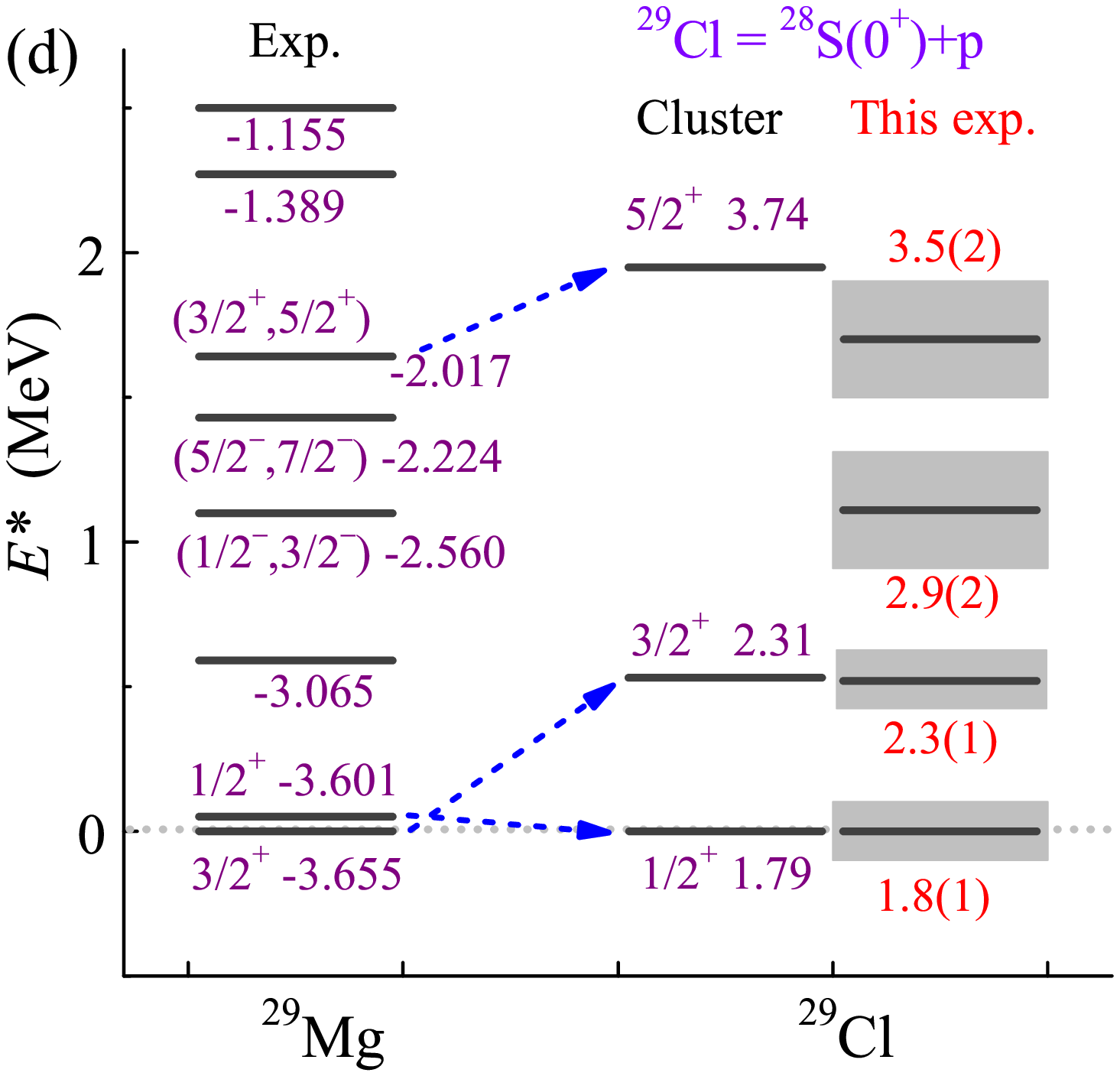}
\includegraphics[width=0.45\textwidth]{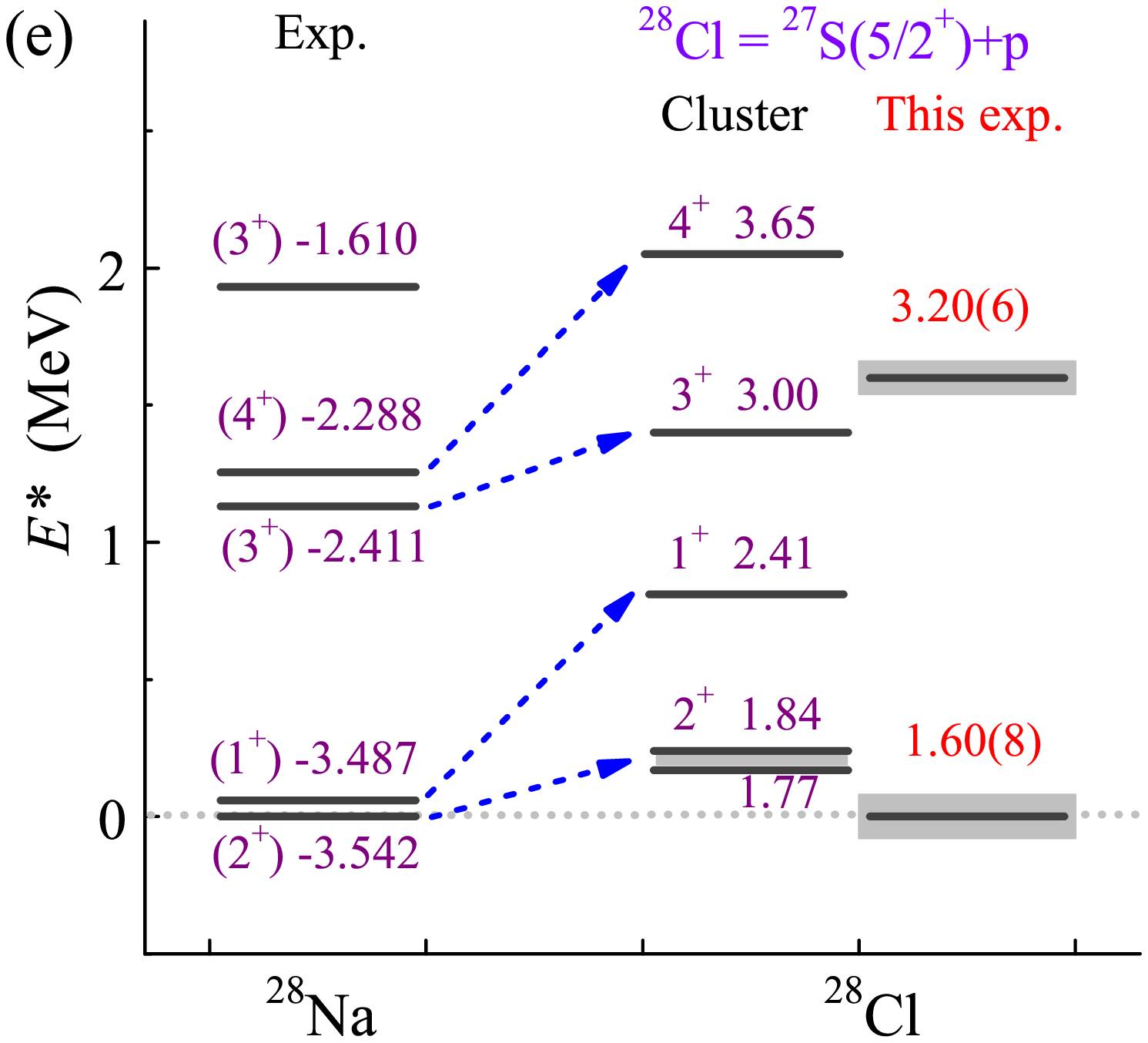}
\end{center}
\caption{  Energy levels of chlorine isotopes. Vertical axes show excitation energies $E^*$. The level legend gives spin-parity $J^{\pi}$ and energies relative to the 1\emph{p}-emission threshold for the chlorine chain members or 1\emph{n}-emission threshold for their isobaric mirror partners.}
\label{fig:levels-cl}
\end{figure*}

A doublet of low-lying states can be found in the bottom of $^{28}$Na spectrum, see Fig.\ \ref{fig:levels-cl} (e). Presumably, the $2^+$ and $1^+$ states are separated by just of $\sim 50$ keV. The $1^+$ state can be only $d$-wave dominated, while $2^+$ can  be either $s$-wave or $d$-wave dominated. If both states have a $d$-wave structure, then the $^{28}$Cl g.s.\ should be found at about 2.4 MeV. In contrast, the observation of decay events corresponding to $E_r=1.60(8)$ MeV can be easily interpreted as the $s$-wave g.s.\ of $^{28}$Cl with the predicted energy of $1.77-1.84$ MeV.

\section{Systematics look on argon isotopes}
\label{sec:sep-cl-ar}

\begin{figure}
\begin{center}
\includegraphics[width=0.46\textwidth]{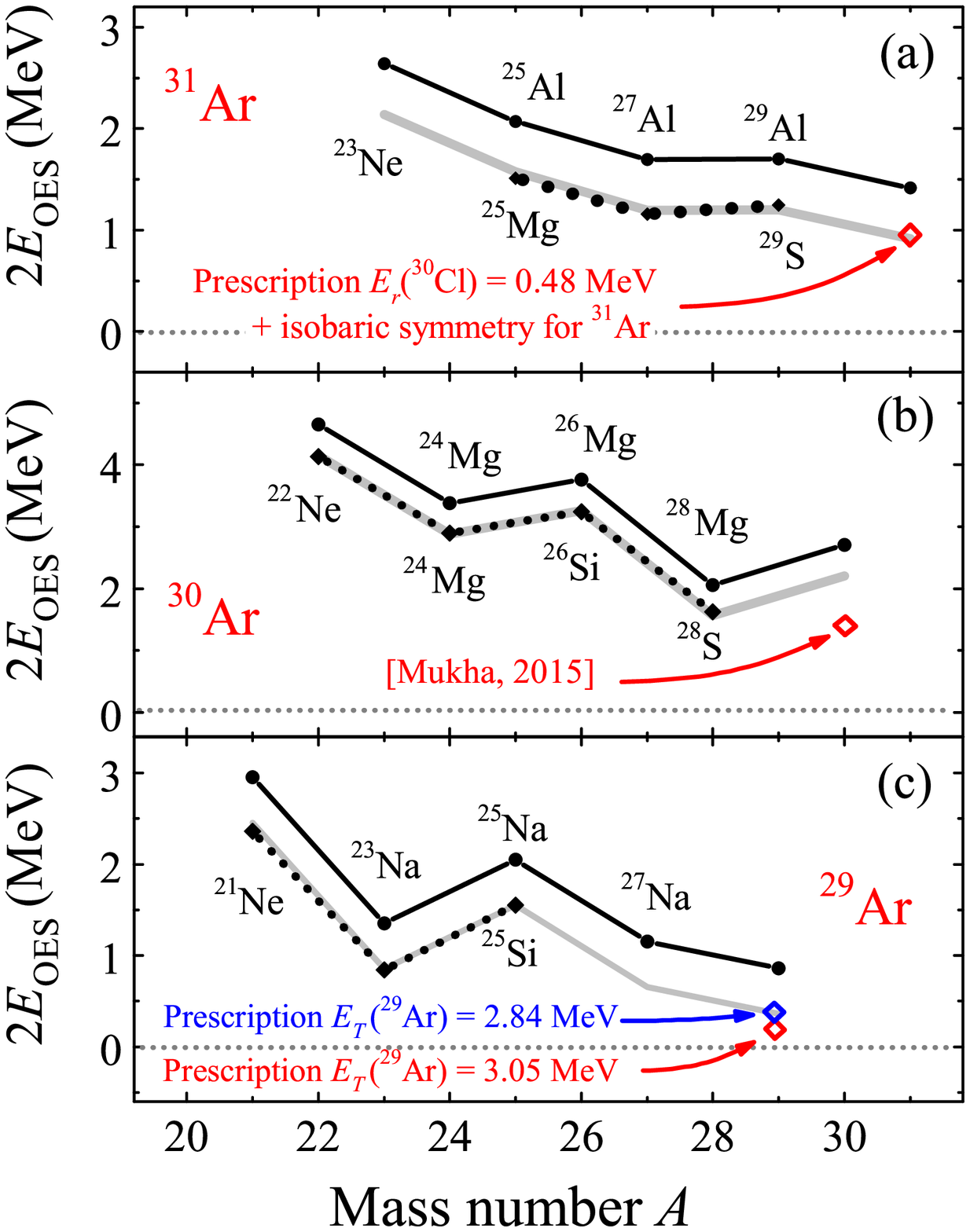}
\end{center}
\caption{  Odd-even staggering (TES) energies for the isotones leading to $^{31}$Ar (a), $^{30}$Ar (b), and $^{29}$Ar (c) are shown by dashed lines. The OES energies for mirror isobar are given by solid lines. Panel (b) corresponds to Ref.\ \cite{Mukha:2015}. The solid gray lines are  provided to guide the eye. They are shifted down by the constant value of 0.5 MeV, which seems to be the very stable value for variety of isobars and isotones. }
\label{fig:oes}
\end{figure}

After we have systematically investigated the behavior of 1\emph{p} separation energies for the chlorine isotopic chain, we can turn to the systematic studies of the corresponding argon isotopic chain, which is based on the obtained information. Namely, we apply the systematics of odd-even staggering  (OES) energies which were shown to be a very helpful indicator concerning the dripline systems in our previous works  \cite{Fomichev:2011,Mukha:2015,Grigorenko:2017}. The OES energy is defined as
\[
2E_{\text{OES}} = S_{2p} - 2S_p \,.
\]
The systematics of $E_{\text{OES}}$ is presented in Fig.\ \ref{fig:oes}. One can see that the systematic trends are very stable for the all  considered isotopic chains. The $E_{\text{OES}}$ is always smaller for the proton-rich systematics compared to the neutron-rich one. The difference of 0.5 MeV is practically the same value for all three cases, see the gray line in Fig.\ \ref{fig:oes}. $E_{\text{OES}}$ also systematically decreases with an increase of mass number, which indicates a borderline of nuclear stability. The $E_{\text{OES}}$ for $^{30}$Ar was found to be smaller than the corresponding systematic expectation \cite{Mukha:2015}. It was argued in this work that such a deviation is typical for systems beyond the dripline, which is confirmed by the examples of well studied 2\emph{p} emitters $^{12}$O, $^{16}$Ne and $^{19}$Mg. Theoretical basis for such an effect is provided by the three-body mechanism of TES \cite{Grigorenko:2002}, which was recently validated by the high-precision data and theoretical calculations in Ref.\ \cite{Grigorenko:2015}. When extrapolating this trend to the nearby isotopes, one may expect that $^{31}$Ar should reside on the $E_{\text{OES}}$ systematics curve or slightly below, while the $^{29}$Ar could be considerably below.

The excitation spectrum of $^{31}$Ar obtained in this work demonstrates a very high level of isobaric symmetry in respect to its mirror $^{31}$Al, see Fig.\ \ref{fig:decay_scheme_31ar}. Based on the isobaric symmetry assumption, we can infer very small value of the 2\emph{p }threshold $S_{2p} = - 3(50)$~keV for the g.s.\ of $^{31}$Ar. This value is obtained by a comparison of the 2\emph{p}-decay energy of 950(50)~keV and the literature value of 946.7(3)~keV of the excitation energy  of the first excited state in $^{31}$Ar and its mirror $^{31}$Al \cite{data:A31_2013}, respectively.
The $S_{2p}$ value of $^{31}$Ar g.s.\ may be also obtained by a comparison of the aligned low-energy exited states in $^{31}$Ar and $^{31}$Al.
Namely, the states in $^{31}$Ar with 2\emph{p}-decay energy of  1.580(60), 2.120(70) and 2.620(130) MeV match the known excited states in $^{31}$Al \cite{data:A31_2013} at excitation energy of  1.613(0.24), 2.090(11) and 2.676(28)~MeV, respectively.
By assuming the same energy between the g.s.\  and the respective excited state both in $^{31}$Ar and $^{31}$Al, we obtain more estimations of the g.s.\ of $^{31}$Ar: at $S_{2p} = {+33(60), -30(81), +56(158)}$~keV, respectively.

The weighted mean of all four pairs provides the averaged $S_{2p}$ value of +6(34)~keV which we finally accept for the g.s.\  of $^{31}$Ar.
Our evaluation agrees within the experimental uncertainties with the previously-estimated $S_{2p}$ value of -3(110)~keV  obtained in beta-decay studies of  $^{31}$Ar \cite{Axelsson:1998}, and precision of the present result is improved by the factor of 3. Our conclusion is that the $^{31}$Ar g.s.\ is  rather bound than not.

With the known value $S_p(^{30}\text{Cl})= -0.48(2)$ MeV, we can estimate the value $2E_{\text{OES}} = 0.966(74)$ MeV for $^{31}$Ar,
which is in a good agreement with the extrapolated OES energy trend in Fig.\ \ref{fig:oes} (a), which gives $2E_{\text{OES}} = 0.915$~MeV. This is an additional argument in favor of the isobaric symmetry (or very close to that) of the $^{31}$Ar and $^{31}$Al ground states.

So, the $^{31}$Ar g.s.\  is evaluated to be likely bound  with the $2p$ separation  energy of less than 40~keV. Even if it is $2p$-unbound (which can not be excluded by our results),  its decay status is not affected: for such a small decay energy the $2p$ partial-lifetime of $^{31}$Ar is incomparably longer than its $\beta$-decay lifetime. Then the g.s.\ of $^{31}$Ar can still be considered as a quasi-stable state in many theoretical applications. An interesting issue here could be the possible existence of the  $2p$-halo  in such an extremely lousy-bound proton-rich nuclide.

Now let us turn to the $^{29}$Ar system. As discussed above, we expect the chlorine isotopes to be overbound in comparison with their mirror isobars (relative to the isobaric symmetry expectations) in a region beyond the dripline because of the TES. For the argon isobars far beyond the dripline, there should be a competition of two trends. One is the overbinding because of TES (the Coulomb displacement energy decreases because of the increase of the valence orbital size). An opposite trend is underbinding due to $E_{\text{OES}}$ reductions (the \emph{p-p} pairing energy decreases because of the increase of the valence-proton orbital size). One must note that the absolute value of extrapolated OES energy is quite low, $2E_{\text{OES}} = 0.361$ MeV [see Fig.\ \ref{fig:oes} (c)]. As  the negative or extremely small value of paring energy seems to be unrealistic assumptions, we accept the following values   $\{E_{\text{OES}},2E_{\text{OES}} \}=\{0.155,0.361 \}$ MeV as the limits of an OES energy variation.
Then we obtain the value $S_{2p} = -2.93(25)$ MeV for the  g.s.\ of $^{29}$Ar by accepting $S_{p} = -1.60(8)$. According to this estimate, the state observed in $^{29}$Ar at $S_{2p} = -5.50(18)$ MeV  can not be assigned as its ground state, and therefore it should be one of the excited states in $^{29}$Ar. However, one should note that this prediction based on the OES systematics is not in accord with the other systematics and the results of theoretical calculations available in the literature, see Table \ref{tab:s2p-sp} reviewing  the published  results on $^{29}$Ar. So, further studies of the $^{29}$Ar system are required in order to clarify the issue.

\begin{table}[htb]
\caption{Separation energies (in MeV) $S_{2p}$ for $^{29}$Ar and  $S_{p}$ for $^{28}$Cl according to different systematics and theoretical
predictions.}
\begin{ruledtabular}
\begin{tabular}[c]{cccccc}
Work  & This, exp. &  This, sys. & \cite{Cole:1998} & \cite{Tian:2013} & \cite{Simonis:2016} \\
\hline
$S_{2p}$  & $-5.50(18)$ & $-2.93(25)$ & $-5.17(16)$ & $-5.17(2)$ & $-8.5(3.5)$ \\
$S_{p}$  & $-1.60(8)$ & $-1.80(4)$ & $-2.84(11)$  & $-2.82(2)$ &   \\
\end{tabular}
\end{ruledtabular}
\label{tab:s2p-sp}
\end{table}

%
%
%


\section{Conclusion}
%

The new isotopes $^{29}$Cl and $^{30}$Ar were recently discovered \cite{Mukha:2015} and the spectroscopy of these two nuclei was performed \cite{Xu:2018} with the reactions of $^{31}$Ar exotic beam at 620~\emph{A}MeV energy on light target. In this work, we investigated the additional inelastic excitation and particle knockout channels of those reactions. The main results of this work are:

\noindent (i) Two previously-unknown isotopes, $^{28}$Cl and $^{30}$Cl, which are unbound respective to the $1p$ emission have been observed. The ground state energies of $^{28}$Cl and $^{30}$Cl have been derived by using angular $^{27,29}$S+$p$ correlations. In addition, four excited states of $^{30}$Cl have been identified as the sub-systems of the previously-unknown excited states of $^{31}$Ar. These states were populated by inelastic excitation of secondary $^{31}$Ar beam and identified by registering $^{29}$S+$p$+$p$ correlations.

\noindent (ii) The first-time observed excitation spectrum of $^{31}$Ar matches very well the excitation spectrum of its isobaric partner $^{31}$Al. The registered isobaric symmetry is used in order to infer the position of the $^{31}$Ar ground state at the \emph{2p} separation energy $S_{2p} = 0.006(34)$ MeV. The high level of isobaric symmetry of these mirror nuclei is confirmed by the systematics of OES energies. The near-zero value of $S_{2p}$ of $^{31}$Ar suggests speculations about the possibility of $2p$ halo in this nuclide.

\noindent (iii) First evidence on a  state in a previously-unobserved isotope $^{29}$Ar has been obtained by detecting $^{27}$S+$p$+$p$ correlations. The state was found to be $2p$-unbound with $S_{2p} = -5.50(18)$ MeV. The results of the different energy systematics do not allow to clarify the status of the observed state. It may be either an excited at $\sim 1-2$ MeV above the ground state (as estimated in this work) or it may be a ground state of $^{29}$Ar according to Refs.~\cite{Cole:1998,Tian:2013,Simonis:2016}. This situation calls for further measurements.


\begin{acknowledgments}
This work was supported in part by the Helmholtz International Center for FAIR (HIC for FAIR), the Helmholtz Association (grant IK-RU-002), the Russian Science Foundation (grant No.\ 17-12-01367), the Polish National Science Center (Contract No.\ UMO-2015/17/B/ST2/00581), the Polish Ministry of Science and Higher Education (Grant No.\ 0079/DIA/2014/43, Grant Diamentowy), the Helmholtz- CAS Joint Research Group (grant HCJRG-108), the FPA2016-77689-C2-1-R contract (MEC, Spain), the MEYS Projects  LTT17003 and LM2015049 (Czech Republic), the Justus-Liebig-Universit\"at Giessen (JLU) and GSI under the JLU-GSI strategic Helmholtz partnership agreement.
\end{acknowledgments}


\bibliographystyle{apsrev4-1}

\bibliography{/u/dkostyl/PRC_paper_Ar/references/all}

\begin{thebibliography}{20}%
\makeatletter
\providecommand \@ifxundefined [1]{%
 \@ifx{#1\undefined}
}%
\providecommand \@ifnum [1]{%
 \ifnum #1\expandafter \@firstoftwo
 \else \expandafter \@secondoftwo
 \fi
}%
\providecommand \@ifx [1]{%
 \ifx #1\expandafter \@firstoftwo
 \else \expandafter \@secondoftwo
 \fi
}%
\providecommand \natexlab [1]{#1}%
\providecommand \enquote  [1]{``#1''}%
\providecommand \bibnamefont  [1]{#1}%
\providecommand \bibfnamefont [1]{#1}%
\providecommand \citenamefont [1]{#1}%
\providecommand \href@noop [0]{\@secondoftwo}%
\providecommand \href [0]{\begingroup \@sanitize@url \@href}%
\providecommand \@href[1]{\@@startlink{#1}\@@href}%
\providecommand \@@href[1]{\endgroup#1\@@endlink}%
\providecommand \@sanitize@url [0]{\catcode `\\12\catcode `\$12\catcode
  `\&12\catcode `\#12\catcode `\^12\catcode `\_12\catcode `\%12\relax}%
\providecommand \@@startlink[1]{}%
\providecommand \@@endlink[0]{}%
\providecommand \url  [0]{\begingroup\@sanitize@url \@url }%
\providecommand \@url [1]{\endgroup\@href {#1}{\urlprefix }}%
\providecommand \urlprefix  [0]{URL }%
\providecommand \Eprint [0]{\href }%
\providecommand \doibase [0]{http://dx.doi.org/}%
\providecommand \selectlanguage [0]{\@gobble}%
\providecommand \bibinfo  [0]{\@secondoftwo}%
\providecommand \bibfield  [0]{\@secondoftwo}%
\providecommand \translation [1]{[#1]}%
\providecommand \BibitemOpen [0]{}%
\providecommand \bibitemStop [0]{}%
\providecommand \bibitemNoStop [0]{.\EOS\space}%
\providecommand \EOS [0]{\spacefactor3000\relax}%
\providecommand \BibitemShut  [1]{\csname bibitem#1\endcsname}%
\let\auto@bib@innerbib\@empty
\bibitem [{\citenamefont {Goldberg}\ \emph {et~al.}(2010)\citenamefont
  {Goldberg}, \citenamefont {Roeder}, \citenamefont {Rogachev}, \citenamefont
  {Chubarian}, \citenamefont {Johnson}, \citenamefont {Fu}, \citenamefont
  {Alharbi}, \citenamefont {Avila}, \citenamefont {Banu}, \citenamefont
  {McCleskey}, \citenamefont {Mitchell}, \citenamefont {Simmons}, \citenamefont
  {Tabacaru}, \citenamefont {Trache},\ and\ \citenamefont
  {Tribble}}]{Goldberg:2010}%
  \BibitemOpen
  \bibfield  {author} {\bibinfo {author} {\bibfnamefont {V.}~\bibnamefont
  {Goldberg}}, \bibinfo {author} {\bibfnamefont {B.}~\bibnamefont {Roeder}},
  \bibinfo {author} {\bibfnamefont {G.}~\bibnamefont {Rogachev}}, \bibinfo
  {author} {\bibfnamefont {G.}~\bibnamefont {Chubarian}}, \bibinfo {author}
  {\bibfnamefont {E.}~\bibnamefont {Johnson}}, \bibinfo {author} {\bibfnamefont
  {C.}~\bibnamefont {Fu}}, \bibinfo {author} {\bibfnamefont {A.}~\bibnamefont
  {Alharbi}}, \bibinfo {author} {\bibfnamefont {M.}~\bibnamefont {Avila}},
  \bibinfo {author} {\bibfnamefont {A.}~\bibnamefont {Banu}}, \bibinfo {author}
  {\bibfnamefont {M.}~\bibnamefont {McCleskey}}, \bibinfo {author}
  {\bibfnamefont {J.}~\bibnamefont {Mitchell}}, \bibinfo {author}
  {\bibfnamefont {E.}~\bibnamefont {Simmons}}, \bibinfo {author} {\bibfnamefont
  {G.}~\bibnamefont {Tabacaru}}, \bibinfo {author} {\bibfnamefont
  {L.}~\bibnamefont {Trache}}, \ and\ \bibinfo {author} {\bibfnamefont
  {R.}~\bibnamefont {Tribble}},\ }\href {\doibase
  https://doi.org/10.1016/j.physletb.2010.07.054} {\bibfield  {journal}
  {\bibinfo  {journal} {Physics Letters B}\ }\textbf {\bibinfo {volume}
  {692}},\ \bibinfo {pages} {307 } (\bibinfo {year} {2010})}\BibitemShut
  {NoStop}%
\bibitem [{\citenamefont {Mukha}\ \emph {et~al.}(2015)\citenamefont {Mukha},
  \citenamefont {Grigorenko}, \citenamefont {Xu}, \citenamefont {Acosta},
  \citenamefont {Casarejos}, \citenamefont {Ciemny}, \citenamefont {Dominik},
  \citenamefont {Du\'enas-D\'{\i}az}, \citenamefont {Dunin}, \citenamefont
  {Espino}, \citenamefont {Estrad\'e}, \citenamefont {Farinon}, \citenamefont
  {Fomichev}, \citenamefont {Geissel}, \citenamefont {Golubkova}, \citenamefont
  {Gorshkov}, \citenamefont {Janas}, \citenamefont {Kami\ifmmode~\acute{n}\else
  \'{n}\fi{}ski}, \citenamefont {Kiselev}, \citenamefont {Kn\"obel},
  \citenamefont {Krupko}, \citenamefont {Kuich}, \citenamefont {Litvinov},
  \citenamefont {Marquinez-Dur\'an}, \citenamefont {Martel}, \citenamefont
  {Mazzocchi}, \citenamefont {Nociforo}, \citenamefont {Ord\'uz}, \citenamefont
  {Pf\"utzner}, \citenamefont {Pietri}, \citenamefont {Pomorski}, \citenamefont
  {Prochazka}, \citenamefont {Rymzhanova}, \citenamefont
  {S\'anchez-Ben\'{\i}tez}, \citenamefont {Scheidenberger}, \citenamefont
  {Sharov}, \citenamefont {Simon}, \citenamefont {Sitar}, \citenamefont
  {Slepnev}, \citenamefont {Stanoiu}, \citenamefont {Strmen}, \citenamefont
  {Szarka}, \citenamefont {Takechi}, \citenamefont {Tanaka}, \citenamefont
  {Weick}, \citenamefont {Winkler}, \citenamefont {Winfield},\ and\
  \citenamefont {Zhukov}}]{Mukha:2015}%
  \BibitemOpen
  \bibfield  {author} {\bibinfo {author} {\bibfnamefont {I.}~\bibnamefont
  {Mukha}}, \bibinfo {author} {\bibfnamefont {L.~V.}\ \bibnamefont
  {Grigorenko}}, \bibinfo {author} {\bibfnamefont {X.}~\bibnamefont {Xu}},
  \bibinfo {author} {\bibfnamefont {L.}~\bibnamefont {Acosta}}, \bibinfo
  {author} {\bibfnamefont {E.}~\bibnamefont {Casarejos}}, \bibinfo {author}
  {\bibfnamefont {A.~A.}\ \bibnamefont {Ciemny}}, \bibinfo {author}
  {\bibfnamefont {W.}~\bibnamefont {Dominik}}, \bibinfo {author} {\bibfnamefont
  {J.}~\bibnamefont {Du\'enas-D\'{\i}az}}, \bibinfo {author} {\bibfnamefont
  {V.}~\bibnamefont {Dunin}}, \bibinfo {author} {\bibfnamefont {J.~M.}\
  \bibnamefont {Espino}}, \bibinfo {author} {\bibfnamefont {A.}~\bibnamefont
  {Estrad\'e}}, \bibinfo {author} {\bibfnamefont {F.}~\bibnamefont {Farinon}},
  \bibinfo {author} {\bibfnamefont {A.}~\bibnamefont {Fomichev}}, \bibinfo
  {author} {\bibfnamefont {H.}~\bibnamefont {Geissel}}, \bibinfo {author}
  {\bibfnamefont {T.~A.}\ \bibnamefont {Golubkova}}, \bibinfo {author}
  {\bibfnamefont {A.}~\bibnamefont {Gorshkov}}, \bibinfo {author}
  {\bibfnamefont {Z.}~\bibnamefont {Janas}}, \bibinfo {author} {\bibfnamefont
  {G.}~\bibnamefont {Kami\ifmmode~\acute{n}\else \'{n}\fi{}ski}}, \bibinfo
  {author} {\bibfnamefont {O.}~\bibnamefont {Kiselev}}, \bibinfo {author}
  {\bibfnamefont {R.}~\bibnamefont {Kn\"obel}}, \bibinfo {author}
  {\bibfnamefont {S.}~\bibnamefont {Krupko}}, \bibinfo {author} {\bibfnamefont
  {M.}~\bibnamefont {Kuich}}, \bibinfo {author} {\bibfnamefont {Y.~A.}\
  \bibnamefont {Litvinov}}, \bibinfo {author} {\bibfnamefont {G.}~\bibnamefont
  {Marquinez-Dur\'an}}, \bibinfo {author} {\bibfnamefont {I.}~\bibnamefont
  {Martel}}, \bibinfo {author} {\bibfnamefont {C.}~\bibnamefont {Mazzocchi}},
  \bibinfo {author} {\bibfnamefont {C.}~\bibnamefont {Nociforo}}, \bibinfo
  {author} {\bibfnamefont {A.~K.}\ \bibnamefont {Ord\'uz}}, \bibinfo {author}
  {\bibfnamefont {M.}~\bibnamefont {Pf\"utzner}}, \bibinfo {author}
  {\bibfnamefont {S.}~\bibnamefont {Pietri}}, \bibinfo {author} {\bibfnamefont
  {M.}~\bibnamefont {Pomorski}}, \bibinfo {author} {\bibfnamefont
  {A.}~\bibnamefont {Prochazka}}, \bibinfo {author} {\bibfnamefont
  {S.}~\bibnamefont {Rymzhanova}}, \bibinfo {author} {\bibfnamefont {A.~M.}\
  \bibnamefont {S\'anchez-Ben\'{\i}tez}}, \bibinfo {author} {\bibfnamefont
  {C.}~\bibnamefont {Scheidenberger}}, \bibinfo {author} {\bibfnamefont
  {P.}~\bibnamefont {Sharov}}, \bibinfo {author} {\bibfnamefont
  {H.}~\bibnamefont {Simon}}, \bibinfo {author} {\bibfnamefont
  {B.}~\bibnamefont {Sitar}}, \bibinfo {author} {\bibfnamefont
  {R.}~\bibnamefont {Slepnev}}, \bibinfo {author} {\bibfnamefont
  {M.}~\bibnamefont {Stanoiu}}, \bibinfo {author} {\bibfnamefont
  {P.}~\bibnamefont {Strmen}}, \bibinfo {author} {\bibfnamefont
  {I.}~\bibnamefont {Szarka}}, \bibinfo {author} {\bibfnamefont
  {M.}~\bibnamefont {Takechi}}, \bibinfo {author} {\bibfnamefont {Y.~K.}\
  \bibnamefont {Tanaka}}, \bibinfo {author} {\bibfnamefont {H.}~\bibnamefont
  {Weick}}, \bibinfo {author} {\bibfnamefont {M.}~\bibnamefont {Winkler}},
  \bibinfo {author} {\bibfnamefont {J.~S.}\ \bibnamefont {Winfield}}, \ and\
  \bibinfo {author} {\bibfnamefont {M.~V.}\ \bibnamefont {Zhukov}},\ }\href
  {\doibase 10.1103/PhysRevLett.115.202501} {\bibfield  {journal} {\bibinfo
  {journal} {Phys. Rev. Lett.}\ }\textbf {\bibinfo {volume} {115}},\ \bibinfo
  {pages} {202501} (\bibinfo {year} {2015})}\BibitemShut {NoStop}%
\bibitem [{\citenamefont {Golubkova}\ \emph {et~al.}(2016)\citenamefont
  {Golubkova}, \citenamefont {Xu}, \citenamefont {Grigorenko}, \citenamefont
  {Mukha}, \citenamefont {Scheidenberger},\ and\ \citenamefont
  {Zhukov}}]{Golubkova:2016}%
  \BibitemOpen
  \bibfield  {author} {\bibinfo {author} {\bibfnamefont {T.}~\bibnamefont
  {Golubkova}}, \bibinfo {author} {\bibfnamefont {X.-D.}\ \bibnamefont {Xu}},
  \bibinfo {author} {\bibfnamefont {L.}~\bibnamefont {Grigorenko}}, \bibinfo
  {author} {\bibfnamefont {I.}~\bibnamefont {Mukha}}, \bibinfo {author}
  {\bibfnamefont {C.}~\bibnamefont {Scheidenberger}}, \ and\ \bibinfo {author}
  {\bibfnamefont {M.}~\bibnamefont {Zhukov}},\ }\href {\doibase
  https://doi.org/10.1016/j.physletb.2016.09.034} {\bibfield  {journal}
  {\bibinfo  {journal} {Physics Letters B}\ }\textbf {\bibinfo {volume}
  {762}},\ \bibinfo {pages} {263 } (\bibinfo {year} {2016})}\BibitemShut
  {NoStop}%
\bibitem [{\citenamefont {Xu}\ \emph {et~al.}(2018)\citenamefont {Xu},
  \citenamefont {Mukha}, \citenamefont {Grigorenko}, \citenamefont
  {Scheidenberger}, \citenamefont {Acosta}, \citenamefont {Casarejos},
  \citenamefont {Chudoba}, \citenamefont {Ciemny}, \citenamefont {Dominik},
  \citenamefont {Du\'enas-D\'{\i}az}, \citenamefont {Dunin}, \citenamefont
  {Espino}, \citenamefont {Estrad\'e}, \citenamefont {Farinon}, \citenamefont
  {Fomichev}, \citenamefont {Geissel}, \citenamefont {Golubkova}, \citenamefont
  {Gorshkov}, \citenamefont {Janas}, \citenamefont {Kami\ifmmode~\acute{n}\else
  \'{n}\fi{}ski}, \citenamefont {Kiselev}, \citenamefont {Kn\"obel},
  \citenamefont {Krupko}, \citenamefont {Kuich}, \citenamefont {Litvinov},
  \citenamefont {Marquinez-Dur\'an}, \citenamefont {Martel}, \citenamefont
  {Mazzocchi}, \citenamefont {Nociforo}, \citenamefont {Ord\'uz}, \citenamefont
  {Pf\"utzner}, \citenamefont {Pietri}, \citenamefont {Pomorski}, \citenamefont
  {Prochazka}, \citenamefont {Rymzhanova}, \citenamefont
  {S\'anchez-Ben\'{\i}tez}, \citenamefont {Sharov}, \citenamefont {Simon},
  \citenamefont {Sitar}, \citenamefont {Slepnev}, \citenamefont {Stanoiu},
  \citenamefont {Strmen}, \citenamefont {Szarka}, \citenamefont {Takechi},
  \citenamefont {Tanaka}, \citenamefont {Weick}, \citenamefont {Winkler},\ and\
  \citenamefont {Winfield}}]{Xu:2018}%
  \BibitemOpen
  \bibfield  {author} {\bibinfo {author} {\bibfnamefont {X.-D.}\ \bibnamefont
  {Xu}}, \bibinfo {author} {\bibfnamefont {I.}~\bibnamefont {Mukha}}, \bibinfo
  {author} {\bibfnamefont {L.~V.}\ \bibnamefont {Grigorenko}}, \bibinfo
  {author} {\bibfnamefont {C.}~\bibnamefont {Scheidenberger}}, \bibinfo
  {author} {\bibfnamefont {L.}~\bibnamefont {Acosta}}, \bibinfo {author}
  {\bibfnamefont {E.}~\bibnamefont {Casarejos}}, \bibinfo {author}
  {\bibfnamefont {V.}~\bibnamefont {Chudoba}}, \bibinfo {author} {\bibfnamefont
  {A.~A.}\ \bibnamefont {Ciemny}}, \bibinfo {author} {\bibfnamefont
  {W.}~\bibnamefont {Dominik}}, \bibinfo {author} {\bibfnamefont
  {J.}~\bibnamefont {Du\'enas-D\'{\i}az}}, \bibinfo {author} {\bibfnamefont
  {V.}~\bibnamefont {Dunin}}, \bibinfo {author} {\bibfnamefont {J.~M.}\
  \bibnamefont {Espino}}, \bibinfo {author} {\bibfnamefont {A.}~\bibnamefont
  {Estrad\'e}}, \bibinfo {author} {\bibfnamefont {F.}~\bibnamefont {Farinon}},
  \bibinfo {author} {\bibfnamefont {A.}~\bibnamefont {Fomichev}}, \bibinfo
  {author} {\bibfnamefont {H.}~\bibnamefont {Geissel}}, \bibinfo {author}
  {\bibfnamefont {T.~A.}\ \bibnamefont {Golubkova}}, \bibinfo {author}
  {\bibfnamefont {A.}~\bibnamefont {Gorshkov}}, \bibinfo {author}
  {\bibfnamefont {Z.}~\bibnamefont {Janas}}, \bibinfo {author} {\bibfnamefont
  {G.}~\bibnamefont {Kami\ifmmode~\acute{n}\else \'{n}\fi{}ski}}, \bibinfo
  {author} {\bibfnamefont {O.}~\bibnamefont {Kiselev}}, \bibinfo {author}
  {\bibfnamefont {R.}~\bibnamefont {Kn\"obel}}, \bibinfo {author}
  {\bibfnamefont {S.}~\bibnamefont {Krupko}}, \bibinfo {author} {\bibfnamefont
  {M.}~\bibnamefont {Kuich}}, \bibinfo {author} {\bibfnamefont {Y.~A.}\
  \bibnamefont {Litvinov}}, \bibinfo {author} {\bibfnamefont {G.}~\bibnamefont
  {Marquinez-Dur\'an}}, \bibinfo {author} {\bibfnamefont {I.}~\bibnamefont
  {Martel}}, \bibinfo {author} {\bibfnamefont {C.}~\bibnamefont {Mazzocchi}},
  \bibinfo {author} {\bibfnamefont {C.}~\bibnamefont {Nociforo}}, \bibinfo
  {author} {\bibfnamefont {A.~K.}\ \bibnamefont {Ord\'uz}}, \bibinfo {author}
  {\bibfnamefont {M.}~\bibnamefont {Pf\"utzner}}, \bibinfo {author}
  {\bibfnamefont {S.}~\bibnamefont {Pietri}}, \bibinfo {author} {\bibfnamefont
  {M.}~\bibnamefont {Pomorski}}, \bibinfo {author} {\bibfnamefont
  {A.}~\bibnamefont {Prochazka}}, \bibinfo {author} {\bibfnamefont
  {S.}~\bibnamefont {Rymzhanova}}, \bibinfo {author} {\bibfnamefont {A.~M.}\
  \bibnamefont {S\'anchez-Ben\'{\i}tez}}, \bibinfo {author} {\bibfnamefont
  {P.}~\bibnamefont {Sharov}}, \bibinfo {author} {\bibfnamefont
  {H.}~\bibnamefont {Simon}}, \bibinfo {author} {\bibfnamefont
  {B.}~\bibnamefont {Sitar}}, \bibinfo {author} {\bibfnamefont
  {R.}~\bibnamefont {Slepnev}}, \bibinfo {author} {\bibfnamefont
  {M.}~\bibnamefont {Stanoiu}}, \bibinfo {author} {\bibfnamefont
  {P.}~\bibnamefont {Strmen}}, \bibinfo {author} {\bibfnamefont
  {I.}~\bibnamefont {Szarka}}, \bibinfo {author} {\bibfnamefont
  {M.}~\bibnamefont {Takechi}}, \bibinfo {author} {\bibfnamefont {Y.~K.}\
  \bibnamefont {Tanaka}}, \bibinfo {author} {\bibfnamefont {H.}~\bibnamefont
  {Weick}}, \bibinfo {author} {\bibfnamefont {M.}~\bibnamefont {Winkler}}, \
  and\ \bibinfo {author} {\bibfnamefont {J.~S.}\ \bibnamefont {Winfield}},\
  }\href {\doibase 10.1103/PhysRevC.97.034305} {\bibfield  {journal} {\bibinfo
  {journal} {Phys. Rev. C}\ }\textbf {\bibinfo {volume} {97}},\ \bibinfo
  {pages} {034305} (\bibinfo {year} {2018})}\BibitemShut {NoStop}%
\bibitem [{\citenamefont {Lis}\ \emph {et~al.}(2015)\citenamefont {Lis},
  \citenamefont {Mazzocchi}, \citenamefont {Dominik}, \citenamefont {Janas},
  \citenamefont {Pf\"utzner}, \citenamefont {Pomorski}, \citenamefont {Acosta},
  \citenamefont {Baraeva}, \citenamefont {Casarejos}, \citenamefont
  {Du\'enas-D\'{\i}az}, \citenamefont {Dunin}, \citenamefont {Espino},
  \citenamefont {Estrade}, \citenamefont {Farinon}, \citenamefont {Fomichev},
  \citenamefont {Geissel}, \citenamefont {Gorshkov}, \citenamefont
  {Kami\ifmmode~\acute{n}\else \'{n}\fi{}ski}, \citenamefont {Kiselev},
  \citenamefont {Kn\"obel}, \citenamefont {Krupko}, \citenamefont {Kuich},
  \citenamefont {Litvinov}, \citenamefont {Marquinez-Dur\'an}, \citenamefont
  {Martel}, \citenamefont {Mukha}, \citenamefont {Nociforo}, \citenamefont
  {Ord\'uz}, \citenamefont {Pietri}, \citenamefont {Prochazka}, \citenamefont
  {S\'anchez-Ben\'{\i}tez}, \citenamefont {Simon}, \citenamefont {Sitar},
  \citenamefont {Slepnev}, \citenamefont {Stanoiu}, \citenamefont {Strmen},
  \citenamefont {Szarka}, \citenamefont {Takechi}, \citenamefont {Tanaka},
  \citenamefont {Weick},\ and\ \citenamefont {Winfield}}]{Lis:2015}%
  \BibitemOpen
  \bibfield  {author} {\bibinfo {author} {\bibfnamefont {A.~A.}\ \bibnamefont
  {Lis}}, \bibinfo {author} {\bibfnamefont {C.}~\bibnamefont {Mazzocchi}},
  \bibinfo {author} {\bibfnamefont {W.}~\bibnamefont {Dominik}}, \bibinfo
  {author} {\bibfnamefont {Z.}~\bibnamefont {Janas}}, \bibinfo {author}
  {\bibfnamefont {M.}~\bibnamefont {Pf\"utzner}}, \bibinfo {author}
  {\bibfnamefont {M.}~\bibnamefont {Pomorski}}, \bibinfo {author}
  {\bibfnamefont {L.}~\bibnamefont {Acosta}}, \bibinfo {author} {\bibfnamefont
  {S.}~\bibnamefont {Baraeva}}, \bibinfo {author} {\bibfnamefont
  {E.}~\bibnamefont {Casarejos}}, \bibinfo {author} {\bibfnamefont
  {J.}~\bibnamefont {Du\'enas-D\'{\i}az}}, \bibinfo {author} {\bibfnamefont
  {V.}~\bibnamefont {Dunin}}, \bibinfo {author} {\bibfnamefont {J.~M.}\
  \bibnamefont {Espino}}, \bibinfo {author} {\bibfnamefont {A.}~\bibnamefont
  {Estrade}}, \bibinfo {author} {\bibfnamefont {F.}~\bibnamefont {Farinon}},
  \bibinfo {author} {\bibfnamefont {A.}~\bibnamefont {Fomichev}}, \bibinfo
  {author} {\bibfnamefont {H.}~\bibnamefont {Geissel}}, \bibinfo {author}
  {\bibfnamefont {A.}~\bibnamefont {Gorshkov}}, \bibinfo {author}
  {\bibfnamefont {G.}~\bibnamefont {Kami\ifmmode~\acute{n}\else
  \'{n}\fi{}ski}}, \bibinfo {author} {\bibfnamefont {O.}~\bibnamefont
  {Kiselev}}, \bibinfo {author} {\bibfnamefont {R.}~\bibnamefont {Kn\"obel}},
  \bibinfo {author} {\bibfnamefont {S.}~\bibnamefont {Krupko}}, \bibinfo
  {author} {\bibfnamefont {M.}~\bibnamefont {Kuich}}, \bibinfo {author}
  {\bibfnamefont {Y.~A.}\ \bibnamefont {Litvinov}}, \bibinfo {author}
  {\bibfnamefont {G.}~\bibnamefont {Marquinez-Dur\'an}}, \bibinfo {author}
  {\bibfnamefont {I.}~\bibnamefont {Martel}}, \bibinfo {author} {\bibfnamefont
  {I.}~\bibnamefont {Mukha}}, \bibinfo {author} {\bibfnamefont
  {C.}~\bibnamefont {Nociforo}}, \bibinfo {author} {\bibfnamefont {A.~K.}\
  \bibnamefont {Ord\'uz}}, \bibinfo {author} {\bibfnamefont {S.}~\bibnamefont
  {Pietri}}, \bibinfo {author} {\bibfnamefont {A.}~\bibnamefont {Prochazka}},
  \bibinfo {author} {\bibfnamefont {A.~M.}\ \bibnamefont
  {S\'anchez-Ben\'{\i}tez}}, \bibinfo {author} {\bibfnamefont {H.}~\bibnamefont
  {Simon}}, \bibinfo {author} {\bibfnamefont {B.}~\bibnamefont {Sitar}},
  \bibinfo {author} {\bibfnamefont {R.}~\bibnamefont {Slepnev}}, \bibinfo
  {author} {\bibfnamefont {M.}~\bibnamefont {Stanoiu}}, \bibinfo {author}
  {\bibfnamefont {P.}~\bibnamefont {Strmen}}, \bibinfo {author} {\bibfnamefont
  {I.}~\bibnamefont {Szarka}}, \bibinfo {author} {\bibfnamefont
  {M.}~\bibnamefont {Takechi}}, \bibinfo {author} {\bibfnamefont
  {Y.}~\bibnamefont {Tanaka}}, \bibinfo {author} {\bibfnamefont
  {H.}~\bibnamefont {Weick}}, \ and\ \bibinfo {author} {\bibfnamefont {J.~S.}\
  \bibnamefont {Winfield}},\ }\href {\doibase 10.1103/PhysRevC.91.064309}
  {\bibfield  {journal} {\bibinfo  {journal} {Phys. Rev. C}\ }\textbf {\bibinfo
  {volume} {91}},\ \bibinfo {pages} {064309} (\bibinfo {year}
  {2015})}\BibitemShut {NoStop}%
\bibitem [{\citenamefont {Stanoiu}\ \emph {et~al.}(2008)\citenamefont
  {Stanoiu}, \citenamefont {Summerer}, \citenamefont {Mukha}, \citenamefont
  {Chatillon}, \citenamefont {Gil}, \citenamefont {Heil}, \citenamefont
  {Hoffman}, \citenamefont {Kiselev}, \citenamefont {Kurz},\ and\ \citenamefont
  {Ott}}]{Stanoiu:2008}%
  \BibitemOpen
  \bibfield  {author} {\bibinfo {author} {\bibfnamefont {M.}~\bibnamefont
  {Stanoiu}}, \bibinfo {author} {\bibfnamefont {K.}~\bibnamefont {Summerer}},
  \bibinfo {author} {\bibfnamefont {I.}~\bibnamefont {Mukha}}, \bibinfo
  {author} {\bibfnamefont {A.}~\bibnamefont {Chatillon}}, \bibinfo {author}
  {\bibfnamefont {E.~C.}\ \bibnamefont {Gil}}, \bibinfo {author} {\bibfnamefont
  {M.}~\bibnamefont {Heil}}, \bibinfo {author} {\bibfnamefont {J.}~\bibnamefont
  {Hoffman}}, \bibinfo {author} {\bibfnamefont {O.}~\bibnamefont {Kiselev}},
  \bibinfo {author} {\bibfnamefont {N.}~\bibnamefont {Kurz}}, \ and\ \bibinfo
  {author} {\bibfnamefont {W.}~\bibnamefont {Ott}},\ }\href {\doibase
  https://doi.org/10.1016/j.nimb.2008.05.116} {\bibfield  {journal} {\bibinfo
  {journal} {Nucl. Instr. Meth. in Phys. Res., B}\ }\textbf {\bibinfo {volume}
  {266}},\ \bibinfo {pages} {4625 } (\bibinfo {year} {2008})}\BibitemShut
  {NoStop}%
\bibitem [{\citenamefont {Mukha}\ \emph {et~al.}(2010)\citenamefont {Mukha},
  \citenamefont {S\"ummerer}, \citenamefont {Acosta}, \citenamefont {Alvarez},
  \citenamefont {Casarejos}, \citenamefont {Chatillon}, \citenamefont
  {Cortina-Gil}, \citenamefont {Egorova}, \citenamefont {Espino}, \citenamefont
  {Fomichev}, \citenamefont {Garc\'ia-Ramos}, \citenamefont {Geissel},
  \citenamefont {G\'omez-Camacho}, \citenamefont {Grigorenko}, \citenamefont
  {Hofmann}, \citenamefont {Kiselev}, \citenamefont {Korsheninnikov},
  \citenamefont {Kurz}, \citenamefont {Litvinov}, \citenamefont {Litvinova},
  \citenamefont {Martel}, \citenamefont {Nociforo}, \citenamefont {Ott},
  \citenamefont {Pf\"utzner}, \citenamefont {Rodr\'iguez-Tajes}, \citenamefont
  {Roeckl}, \citenamefont {Stanoiu}, \citenamefont {Timofeyuk}, \citenamefont
  {Weick},\ and\ \citenamefont {Woods}}]{Mukha:2010}%
  \BibitemOpen
  \bibfield  {author} {\bibinfo {author} {\bibfnamefont {I.}~\bibnamefont
  {Mukha}}, \bibinfo {author} {\bibfnamefont {K.}~\bibnamefont {S\"ummerer}},
  \bibinfo {author} {\bibfnamefont {L.}~\bibnamefont {Acosta}}, \bibinfo
  {author} {\bibfnamefont {M.~A.~G.}\ \bibnamefont {Alvarez}}, \bibinfo
  {author} {\bibfnamefont {E.}~\bibnamefont {Casarejos}}, \bibinfo {author}
  {\bibfnamefont {A.}~\bibnamefont {Chatillon}}, \bibinfo {author}
  {\bibfnamefont {D.}~\bibnamefont {Cortina-Gil}}, \bibinfo {author}
  {\bibfnamefont {I.~A.}\ \bibnamefont {Egorova}}, \bibinfo {author}
  {\bibfnamefont {J.~M.}\ \bibnamefont {Espino}}, \bibinfo {author}
  {\bibfnamefont {A.}~\bibnamefont {Fomichev}}, \bibinfo {author}
  {\bibfnamefont {J.~E.}\ \bibnamefont {Garc\'ia-Ramos}}, \bibinfo {author}
  {\bibfnamefont {H.}~\bibnamefont {Geissel}}, \bibinfo {author} {\bibfnamefont
  {J.}~\bibnamefont {G\'omez-Camacho}}, \bibinfo {author} {\bibfnamefont
  {L.}~\bibnamefont {Grigorenko}}, \bibinfo {author} {\bibfnamefont
  {J.}~\bibnamefont {Hofmann}}, \bibinfo {author} {\bibfnamefont
  {O.}~\bibnamefont {Kiselev}}, \bibinfo {author} {\bibfnamefont
  {A.}~\bibnamefont {Korsheninnikov}}, \bibinfo {author} {\bibfnamefont
  {N.}~\bibnamefont {Kurz}}, \bibinfo {author} {\bibfnamefont {Y.~A.}\
  \bibnamefont {Litvinov}}, \bibinfo {author} {\bibfnamefont {E.}~\bibnamefont
  {Litvinova}}, \bibinfo {author} {\bibfnamefont {I.}~\bibnamefont {Martel}},
  \bibinfo {author} {\bibfnamefont {C.}~\bibnamefont {Nociforo}}, \bibinfo
  {author} {\bibfnamefont {W.}~\bibnamefont {Ott}}, \bibinfo {author}
  {\bibfnamefont {M.}~\bibnamefont {Pf\"utzner}}, \bibinfo {author}
  {\bibfnamefont {C.}~\bibnamefont {Rodr\'iguez-Tajes}}, \bibinfo {author}
  {\bibfnamefont {E.}~\bibnamefont {Roeckl}}, \bibinfo {author} {\bibfnamefont
  {M.}~\bibnamefont {Stanoiu}}, \bibinfo {author} {\bibfnamefont {N.~K.}\
  \bibnamefont {Timofeyuk}}, \bibinfo {author} {\bibfnamefont {H.}~\bibnamefont
  {Weick}}, \ and\ \bibinfo {author} {\bibfnamefont {P.~J.}\ \bibnamefont
  {Woods}},\ }\href {\doibase 10.1103/PhysRevC.82.054315} {\bibfield  {journal}
  {\bibinfo  {journal} {Phys. Rev. C}\ }\textbf {\bibinfo {volume} {82}},\
  \bibinfo {pages} {054315} (\bibinfo {year} {2010})}\BibitemShut {NoStop}%
\bibitem [{\citenamefont {Mukha}\ \emph {et~al.}(2012)\citenamefont {Mukha},
  \citenamefont {Grigorenko}, \citenamefont {Acosta}, \citenamefont {Alvarez},
  \citenamefont {Casarejos}, \citenamefont {Chatillon}, \citenamefont
  {Cortina-Gil}, \citenamefont {Espino}, \citenamefont {Fomichev},
  \citenamefont {Garc\'{i}a-Ramos}, \citenamefont {Geissel}, \citenamefont
  {G\'omez-Camacho}, \citenamefont {Hofmann}, \citenamefont {Kiselev},
  \citenamefont {Korsheninnikov}, \citenamefont {Kurz}, \citenamefont
  {Litvinov}, \citenamefont {Martel}, \citenamefont {Nociforo}, \citenamefont
  {Ott}, \citenamefont {Pf\"utzner}, \citenamefont {Rodr\'{i}guez-Tajes},
  \citenamefont {Roeckl}, \citenamefont {Scheidenberger}, \citenamefont
  {Stanoiu}, \citenamefont {S\"ummerer}, \citenamefont {Weick},\ and\
  \citenamefont {Woods}}]{Mukha:2012}%
  \BibitemOpen
  \bibfield  {author} {\bibinfo {author} {\bibfnamefont {I.}~\bibnamefont
  {Mukha}}, \bibinfo {author} {\bibfnamefont {L.}~\bibnamefont {Grigorenko}},
  \bibinfo {author} {\bibfnamefont {L.}~\bibnamefont {Acosta}}, \bibinfo
  {author} {\bibfnamefont {M.~A.~G.}\ \bibnamefont {Alvarez}}, \bibinfo
  {author} {\bibfnamefont {E.}~\bibnamefont {Casarejos}}, \bibinfo {author}
  {\bibfnamefont {A.}~\bibnamefont {Chatillon}}, \bibinfo {author}
  {\bibfnamefont {D.}~\bibnamefont {Cortina-Gil}}, \bibinfo {author}
  {\bibfnamefont {J.~M.}\ \bibnamefont {Espino}}, \bibinfo {author}
  {\bibfnamefont {A.}~\bibnamefont {Fomichev}}, \bibinfo {author}
  {\bibfnamefont {J.~E.}\ \bibnamefont {Garc\'{i}a-Ramos}}, \bibinfo {author}
  {\bibfnamefont {H.}~\bibnamefont {Geissel}}, \bibinfo {author} {\bibfnamefont
  {J.}~\bibnamefont {G\'omez-Camacho}}, \bibinfo {author} {\bibfnamefont
  {J.}~\bibnamefont {Hofmann}}, \bibinfo {author} {\bibfnamefont
  {O.}~\bibnamefont {Kiselev}}, \bibinfo {author} {\bibfnamefont
  {A.}~\bibnamefont {Korsheninnikov}}, \bibinfo {author} {\bibfnamefont
  {N.}~\bibnamefont {Kurz}}, \bibinfo {author} {\bibfnamefont {Y.~A.}\
  \bibnamefont {Litvinov}}, \bibinfo {author} {\bibfnamefont {I.}~\bibnamefont
  {Martel}}, \bibinfo {author} {\bibfnamefont {C.}~\bibnamefont {Nociforo}},
  \bibinfo {author} {\bibfnamefont {W.}~\bibnamefont {Ott}}, \bibinfo {author}
  {\bibfnamefont {M.}~\bibnamefont {Pf\"utzner}}, \bibinfo {author}
  {\bibfnamefont {C.}~\bibnamefont {Rodr\'{i}guez-Tajes}}, \bibinfo {author}
  {\bibfnamefont {E.}~\bibnamefont {Roeckl}}, \bibinfo {author} {\bibfnamefont
  {C.}~\bibnamefont {Scheidenberger}}, \bibinfo {author} {\bibfnamefont
  {M.}~\bibnamefont {Stanoiu}}, \bibinfo {author} {\bibfnamefont
  {K.}~\bibnamefont {S\"ummerer}}, \bibinfo {author} {\bibfnamefont
  {H.}~\bibnamefont {Weick}}, \ and\ \bibinfo {author} {\bibfnamefont {P.~J.}\
  \bibnamefont {Woods}},\ }\href {\doibase 10.1103/PhysRevC.85.044325}
  {\bibfield  {journal} {\bibinfo  {journal} {Phys. Rev. C}\ }\textbf {\bibinfo
  {volume} {85}},\ \bibinfo {pages} {044325} (\bibinfo {year}
  {2012})}\BibitemShut {NoStop}%
\bibitem [{\citenamefont {Ehrman}(1951)}]{Ehrman:1951}%
  \BibitemOpen
  \bibfield  {author} {\bibinfo {author} {\bibfnamefont {J.~B.}\ \bibnamefont
  {Ehrman}},\ }\href {\doibase 10.1103/PhysRev.81.412} {\bibfield  {journal}
  {\bibinfo  {journal} {Phys. Rev.}\ }\textbf {\bibinfo {volume} {81}},\
  \bibinfo {pages} {412} (\bibinfo {year} {1951})}\BibitemShut {NoStop}%
\bibitem [{\citenamefont {Thomas}(1952)}]{Thomas:1952}%
  \BibitemOpen
  \bibfield  {author} {\bibinfo {author} {\bibfnamefont {R.~G.}\ \bibnamefont
  {Thomas}},\ }\href {\doibase 10.1103/PhysRev.88.1109} {\bibfield  {journal}
  {\bibinfo  {journal} {Phys. Rev.}\ }\textbf {\bibinfo {volume} {88}},\
  \bibinfo {pages} {1109} (\bibinfo {year} {1952})}\BibitemShut {NoStop}%
\bibitem [{\citenamefont {Angeli}\ and\ \citenamefont
  {Marinova}(2013)}]{Angeli:2013}%
  \BibitemOpen
  \bibfield  {author} {\bibinfo {author} {\bibfnamefont {I.}~\bibnamefont
  {Angeli}}\ and\ \bibinfo {author} {\bibfnamefont {K.}~\bibnamefont
  {Marinova}},\ }\href {\doibase http://dx.doi.org/10.1016/j.adt.2011.12.006}
  {\bibfield  {journal} {\bibinfo  {journal} {Atomic Data and Nuclear Data
  Tables}\ }\textbf {\bibinfo {volume} {99}},\ \bibinfo {pages} {69 } (\bibinfo
  {year} {2013})}\BibitemShut {NoStop}%
\bibitem [{\citenamefont {Fomichev}\ \emph {et~al.}(2011)\citenamefont
  {Fomichev}, \citenamefont {Mukha}, \citenamefont {Stepantsov}, \citenamefont
  {Grigorenko}, \citenamefont {Litvinova}, \citenamefont {Chudoba},
  \citenamefont {Egorova}, \citenamefont {Golovkov}, \citenamefont {Gorshkov},
  \citenamefont {Gorshkov}, \citenamefont {Kaminski}, \citenamefont {Krupko},
  \citenamefont {Parfenova}, \citenamefont {Sidorchuk}, \citenamefont
  {Slepnev}, \citenamefont {Ter-Akopian}, \citenamefont {Wolski},\ and\
  \citenamefont {Zhukov}}]{Fomichev:2011}%
  \BibitemOpen
  \bibfield  {author} {\bibinfo {author} {\bibfnamefont {A.~S.}\ \bibnamefont
  {Fomichev}}, \bibinfo {author} {\bibfnamefont {I.~G.}\ \bibnamefont {Mukha}},
  \bibinfo {author} {\bibfnamefont {S.~V.}\ \bibnamefont {Stepantsov}},
  \bibinfo {author} {\bibfnamefont {L.~V.}\ \bibnamefont {Grigorenko}},
  \bibinfo {author} {\bibfnamefont {E.~V.}\ \bibnamefont {Litvinova}}, \bibinfo
  {author} {\bibfnamefont {V.}~\bibnamefont {Chudoba}}, \bibinfo {author}
  {\bibfnamefont {I.~A.}\ \bibnamefont {Egorova}}, \bibinfo {author}
  {\bibfnamefont {M.~S.}\ \bibnamefont {Golovkov}}, \bibinfo {author}
  {\bibfnamefont {A.~V.}\ \bibnamefont {Gorshkov}}, \bibinfo {author}
  {\bibfnamefont {V.~A.}\ \bibnamefont {Gorshkov}}, \bibinfo {author}
  {\bibfnamefont {G.}~\bibnamefont {Kaminski}}, \bibinfo {author}
  {\bibfnamefont {S.~A.}\ \bibnamefont {Krupko}}, \bibinfo {author}
  {\bibfnamefont {Y.~L.}\ \bibnamefont {Parfenova}}, \bibinfo {author}
  {\bibfnamefont {S.~I.}\ \bibnamefont {Sidorchuk}}, \bibinfo {author}
  {\bibfnamefont {R.~S.}\ \bibnamefont {Slepnev}}, \bibinfo {author}
  {\bibfnamefont {G.~M.}\ \bibnamefont {Ter-Akopian}}, \bibinfo {author}
  {\bibfnamefont {R.}~\bibnamefont {Wolski}}, \ and\ \bibinfo {author}
  {\bibfnamefont {M.~V.}\ \bibnamefont {Zhukov}},\ }\href {\doibase
  10.1142/S0218301311018216} {\bibfield  {journal} {\bibinfo  {journal} {Int.
  Journal of Modern Physics E}\ }\textbf {\bibinfo {volume} {20}},\ \bibinfo
  {pages} {1491} (\bibinfo {year} {2011})}\BibitemShut {NoStop}%
\bibitem [{\citenamefont {Grigorenko}\ \emph {et~al.}(2017)\citenamefont
  {Grigorenko}, \citenamefont {Golubkova}, \citenamefont {Vaagen},\ and\
  \citenamefont {Zhukov}}]{Grigorenko:2017}%
  \BibitemOpen
  \bibfield  {author} {\bibinfo {author} {\bibfnamefont {L.~V.}\ \bibnamefont
  {Grigorenko}}, \bibinfo {author} {\bibfnamefont {T.~A.}\ \bibnamefont
  {Golubkova}}, \bibinfo {author} {\bibfnamefont {J.~S.}\ \bibnamefont
  {Vaagen}}, \ and\ \bibinfo {author} {\bibfnamefont {M.~V.}\ \bibnamefont
  {Zhukov}},\ }\href {\doibase 10.1103/PhysRevC.95.021601} {\bibfield
  {journal} {\bibinfo  {journal} {Phys. Rev. C}\ }\textbf {\bibinfo {volume}
  {95}},\ \bibinfo {pages} {021601} (\bibinfo {year} {2017})}\BibitemShut
  {NoStop}%
\bibitem [{\citenamefont {Grigorenko}\ \emph {et~al.}(2002)\citenamefont
  {Grigorenko}, \citenamefont {Mukha}, \citenamefont {Thompson},\ and\
  \citenamefont {Zhukov}}]{Grigorenko:2002}%
  \BibitemOpen
  \bibfield  {author} {\bibinfo {author} {\bibfnamefont {L.~V.}\ \bibnamefont
  {Grigorenko}}, \bibinfo {author} {\bibfnamefont {I.~G.}\ \bibnamefont
  {Mukha}}, \bibinfo {author} {\bibfnamefont {I.~J.}\ \bibnamefont {Thompson}},
  \ and\ \bibinfo {author} {\bibfnamefont {M.~V.}\ \bibnamefont {Zhukov}},\
  }\href {\doibase 10.1103/PhysRevLett.88.042502} {\bibfield  {journal}
  {\bibinfo  {journal} {Phys. Rev. Lett.}\ }\textbf {\bibinfo {volume} {88}},\
  \bibinfo {pages} {042502} (\bibinfo {year} {2002})}\BibitemShut {NoStop}%
\bibitem [{\citenamefont {Grigorenko}\ \emph {et~al.}(2015)\citenamefont
  {Grigorenko}, \citenamefont {Golubkova},\ and\ \citenamefont
  {Zhukov}}]{Grigorenko:2015}%
  \BibitemOpen
  \bibfield  {author} {\bibinfo {author} {\bibfnamefont {L.~V.}\ \bibnamefont
  {Grigorenko}}, \bibinfo {author} {\bibfnamefont {T.~A.}\ \bibnamefont
  {Golubkova}}, \ and\ \bibinfo {author} {\bibfnamefont {M.~V.}\ \bibnamefont
  {Zhukov}},\ }\href {\doibase 10.1103/PhysRevC.91.024325} {\bibfield
  {journal} {\bibinfo  {journal} {Phys. Rev. C}\ }\textbf {\bibinfo {volume}
  {91}},\ \bibinfo {pages} {024325} (\bibinfo {year} {2015})}\BibitemShut
  {NoStop}%
\bibitem [{\citenamefont {Ouellet}\ and\ \citenamefont
  {Singh}(2013)}]{data:A31_2013}%
  \BibitemOpen
  \bibfield  {author} {\bibinfo {author} {\bibfnamefont {C.}~\bibnamefont
  {Ouellet}}\ and\ \bibinfo {author} {\bibfnamefont {B.}~\bibnamefont
  {Singh}},\ }\href {\doibase https://doi.org/10.1016/j.nds.2013.03.001}
  {\bibfield  {journal} {\bibinfo  {journal} {Nuclear Data Sheets}\ }\textbf
  {\bibinfo {volume} {114}},\ \bibinfo {pages} {209 } (\bibinfo {year}
  {2013})}\BibitemShut {NoStop}%
\bibitem [{\citenamefont {Axelsson}\ \emph {et~al.}(1998)\citenamefont
  {Axelsson}, \citenamefont {Aysto}, \citenamefont {Bergmann}, \citenamefont
  {Borge}, \citenamefont {Fraile}, \citenamefont {Fynbo}, \citenamefont
  {Honkanen}, \citenamefont {Hornshoj}, \citenamefont {Jokinen}, \citenamefont
  {Jonson}, \citenamefont {Martel}, \citenamefont {Mukha}, \citenamefont
  {Nilsson}, \citenamefont {Nyman}, \citenamefont {Petersen}, \citenamefont
  {Riisager}, \citenamefont {Smedberg},\ and\ \citenamefont
  {Tengblad}}]{Axelsson:1998}%
  \BibitemOpen
  \bibfield  {author} {\bibinfo {author} {\bibfnamefont {L.}~\bibnamefont
  {Axelsson}}, \bibinfo {author} {\bibfnamefont {J.}~\bibnamefont {Aysto}},
  \bibinfo {author} {\bibfnamefont {U.}~\bibnamefont {Bergmann}}, \bibinfo
  {author} {\bibfnamefont {M.}~\bibnamefont {Borge}}, \bibinfo {author}
  {\bibfnamefont {L.}~\bibnamefont {Fraile}}, \bibinfo {author} {\bibfnamefont
  {H.}~\bibnamefont {Fynbo}}, \bibinfo {author} {\bibfnamefont
  {A.}~\bibnamefont {Honkanen}}, \bibinfo {author} {\bibfnamefont
  {P.}~\bibnamefont {Hornshoj}}, \bibinfo {author} {\bibfnamefont
  {A.}~\bibnamefont {Jokinen}}, \bibinfo {author} {\bibfnamefont
  {B.}~\bibnamefont {Jonson}}, \bibinfo {author} {\bibfnamefont
  {I.}~\bibnamefont {Martel}}, \bibinfo {author} {\bibfnamefont
  {I.}~\bibnamefont {Mukha}}, \bibinfo {author} {\bibfnamefont
  {T.}~\bibnamefont {Nilsson}}, \bibinfo {author} {\bibfnamefont
  {G.}~\bibnamefont {Nyman}}, \bibinfo {author} {\bibfnamefont
  {B.}~\bibnamefont {Petersen}}, \bibinfo {author} {\bibfnamefont
  {K.}~\bibnamefont {Riisager}}, \bibinfo {author} {\bibfnamefont
  {M.}~\bibnamefont {Smedberg}}, \ and\ \bibinfo {author} {\bibfnamefont
  {O.}~\bibnamefont {Tengblad}},\ }\href {\doibase
  https://doi.org/10.1016/S0375-9474(97)00623-4} {\bibfield  {journal}
  {\bibinfo  {journal} {Nuclear Physics A}\ }\textbf {\bibinfo {volume}
  {628}},\ \bibinfo {pages} {345 } (\bibinfo {year} {1998})}\BibitemShut
  {NoStop}%
\bibitem [{\citenamefont {Cole}(1998)}]{Cole:1998}%
  \BibitemOpen
  \bibfield  {author} {\bibinfo {author} {\bibfnamefont {B.~J.}\ \bibnamefont
  {Cole}},\ }\href {\doibase 10.1103/PhysRevC.58.2831} {\bibfield  {journal}
  {\bibinfo  {journal} {Phys. Rev. C}\ }\textbf {\bibinfo {volume} {58}},\
  \bibinfo {pages} {2831} (\bibinfo {year} {1998})}\BibitemShut {NoStop}%
\bibitem [{\citenamefont {Tian}\ \emph {et~al.}(2013)\citenamefont {Tian},
  \citenamefont {Wang}, \citenamefont {Li},\ and\ \citenamefont
  {Li}}]{Tian:2013}%
  \BibitemOpen
  \bibfield  {author} {\bibinfo {author} {\bibfnamefont {J.}~\bibnamefont
  {Tian}}, \bibinfo {author} {\bibfnamefont {N.}~\bibnamefont {Wang}}, \bibinfo
  {author} {\bibfnamefont {C.}~\bibnamefont {Li}}, \ and\ \bibinfo {author}
  {\bibfnamefont {J.}~\bibnamefont {Li}},\ }\href {\doibase
  10.1103/PhysRevC.87.014313} {\bibfield  {journal} {\bibinfo  {journal} {Phys.
  Rev. C}\ }\textbf {\bibinfo {volume} {87}},\ \bibinfo {pages} {014313}
  (\bibinfo {year} {2013})}\BibitemShut {NoStop}%
\bibitem [{\citenamefont {Simonis}\ \emph {et~al.}(2016)\citenamefont
  {Simonis}, \citenamefont {Hebeler}, \citenamefont {Holt}, \citenamefont
  {Men\'endez},\ and\ \citenamefont {Schwenk}}]{Simonis:2016}%
  \BibitemOpen
  \bibfield  {author} {\bibinfo {author} {\bibfnamefont {J.}~\bibnamefont
  {Simonis}}, \bibinfo {author} {\bibfnamefont {K.}~\bibnamefont {Hebeler}},
  \bibinfo {author} {\bibfnamefont {J.~D.}\ \bibnamefont {Holt}}, \bibinfo
  {author} {\bibfnamefont {J.}~\bibnamefont {Men\'endez}}, \ and\ \bibinfo
  {author} {\bibfnamefont {A.}~\bibnamefont {Schwenk}},\ }\href {\doibase
  10.1103/PhysRevC.93.011302} {\bibfield  {journal} {\bibinfo  {journal} {Phys.
  Rev. C}\ }\textbf {\bibinfo {volume} {93}},\ \bibinfo {pages} {011302}
  (\bibinfo {year} {2016})}\BibitemShut {NoStop}%
\end{thebibliography}%


\end{document}